\documentclass[preprintnumbers,prd,nofootinbib]{revtex4}

\usepackage{graphicx}
\usepackage{latexsym}
\usepackage{amsfonts}
\usepackage{amssymb}
\usepackage{amsmath}
\usepackage{slashed}
\usepackage{feynmp}

\begin{document}

\title{LHC Searches for Non-Chiral Weakly Charged Multiplets}
\author{Matthew R. Buckley$^1$, Lisa Randall$^2$, and Brian Shuve$^2$}
\affiliation{$^1$Department of Physics, California Institute of Technology, Pasadena, CA 91125, USA}
\affiliation{$^2$Harvard University, Cambridge, MA 02138, USA}
\date{\today}

\preprint{CALT-68-2749}

\begin{abstract}
Because the TeV-scale to be probed at the Large Hadron Collider
should shed light on the naturalness, hierarchy, and dark matter
problems, most searches to date have focused on new physics
signatures motivated by possible solutions to these puzzles. In this
paper, we consider some candidates for new states that although not
well-motivated from this standpoint are obvious possibilities that
current search strategies would miss. In particular we consider
vector representations of fermions in multiplets of $SU(2)_L$ with a
lightest neutral state.  Standard search strategies would fail to
find such particles because of the expected small one-loop-level
splitting between charged and neutral states.
\end{abstract}


\maketitle

\section{Introduction}

Based on the Higgs mechanism and the hierarchy problem, we expect
new physical states to be produced at the TeV scale. Current LHC
strategies focus on discovery of different models that address these
problems, such as supersymmetry (SUSY) \cite{Wess:1974tw}, extra
dimensions \cite{ArkaniHamed:1998nn,Randall:1999ee} and technicolor
\cite{Weinstein:1973gj,Weinberg:1979bn,Susskind:1979}.

However, other states may appear at the TeV scale that are
irrelevant to the working of the Higgs mechanism or solutions to the
hierarchy problem but might nonetheless be part of a TeV-sector. An
obvious example would be states that are nonchiral under electroweak
gauge symmetry. Such states can be motivated in specific theories --
as dark matter candidates
\cite{Moroi:1999zb,Cirelli:2005uq,Cui:2009xq},
or to relax constraints from electro-weak symmetry breaking (EWSB)
\cite{FileviezPerez:2008bj}. In this paper we will not assume any
specific model-building goal but assume only that as with dark
matter candidates, the lightest state in the multiplets is neutral.

In particular, we consider  new vector-like fermions in multiplets
of $SU(2)_L$ where the splitting between the charged and neutral
states is due solely to loop corrections, which will generally be
the case unless there are special additional interactions. As a
consequence, the spectrum in these theories is nearly degenerate;
only a few hundred MeV separate the states.

Remarkably, current searches would not find such multiplets, even if
they are as light as 100 GeV. This surprising fact is the result of
the small one-loop splittings, which imply  final states with very
soft Standard Model (SM) fields and an invisible uncharged heavy
particle. The charged states decay rapidly into the neutral particle
and light SM fields (typically electrons, neutrinos, or pions) while
the neutral states (which may or may not be a viable thermal DM
candidates) are stable and so escape undetected. The physical
distance traveled by the charged state depends strongly on the
splitting $\Delta M$; it is on the order of 10 meters when $\Delta M
< m_\pi$, dropping to $\sim10$~cm once the two-body pion channel
opens up ($\Delta M \gtrsim m_\pi$), and falling rapidly for larger
mass splittings.

Current search strategies for new long-lived particles rely on
looking for charged states that make it all the way through the
detector \cite{Drees:1990yw,Aaltonen:2009ke}. Such searches will not
find the vector multiplets we are considering; they would find only
those multiplets that contain exclusively charged states or those
that have $\Delta M < m_\pi$, and thus a lifetime long enough to
reach the muon chamber. However, if the lightest state in the
multiplet is neutral and the splitting is bigger, the detector would
register only short `stub' tracks from the charged states. This
means that, even if these particles can be produced at reasonably
large rates, standard searches would miss them.

The case of $SU(2)_L$ triplet scalars has been investigated in
Ref.~\cite{FileviezPerez:2008bj}. Such models have similar
splittings and couplings, and so have a great deal of collider
phenomenology in common with the fermion case considered in this
paper. These scalar fields will also in the most general case have
tree-level couplings to the Higgs, which could be used to
distinguish between the scalar and fermionic models.

Wino lightest supersymmetric particles (LSPs) that are nearly
degenerate with a charged next-to-lightest supersymmetric particle
(NLSP) are an example of a `complete' model which has similar
phenomenology to the simple vector multiplets. As a result, the
experimental considerations are closely related
\cite{Feng:1999fu,Gunion:1999jr,Abreu:1999qr,Ibe:2006de}. Such a
SUSY spectrum occurs in anomaly-mediated supersymmetry breaking
\cite{Randall:1998uk,Giudice:1998xp}, when either $|\mu| \gg M_1 >
M_2$ or $M_{1/2} \gg |\mu|$. In both situations, the LSP is the
neutralino $\tilde{\chi}^0_1$ and the next-to-lightest
supersymmetric particle (NLSP) is the chargino $\tilde{\chi}^\pm_1$.
In the former case both are primarily wino. In this limit the
tree-level splitting is suppressed by $|\mu|^2$, and is commensurate
with the loop splitting common to all $SU(2)_L$ triplets. The latter
case (small $|\mu|$) is very similar, though here the LSP and NLSP
are primarily higgsino. Since only the LSP and NLSP would typically
be accessible in these scenarios, searches for such spectra has long
been recognized as extremely challenging. Proposals for search
strategies have been made for the Tevatron
\cite{Feng:1999fu,Gunion:1999jr}, the LHC \cite{Ibe:2006de}, and
been carried out at DELPHI \cite{Abreu:1999qr} and OPAL
\cite{Abbiendi:2002vz}. As this phenomenology is essentially an
example of a generic degenerate multiplet, the search proposed here
overlaps with this body of previous work. It is important to
recognize that those searches have potentially greater applicability
than just anomaly-mediated supersymmetry breaking.

The searches we consider in this paper would be quite worthwhile in
that current bounds on such states are pitifully weak.  New
particles coupling to the $Z$ boson must be heavier than $m_Z/2$
\cite{Decamp:1989fr}. The LEP-II search by DELPHI
\cite{Abreu:1999qr} placed a lower bound on degenerate winos of
approximately $90$~GeV for mass splittings less then $\sim 200$~MeV.
OPAL excluded a $95$~GeV higgsino- or wino-like LSP/NLSP pair with splittings larger
than $0.5$~GeV \cite{Abbiendi:2002vz}.
The limits are relaxed to $50-60$~GeV for GeV-scale splittings.

Several additional proposals have been made for alternative searches
at colliders in current operation. However, these have not yet been
implemented. For splittings between $300-600$~MeV, the Tevatron
might be able to place limits between $68(95)$ and $53(75)$~GeV with
an integrated luminosity of $2(30)$ fb$^{-1}$ using searches
for short `stubs' of the kind investigated in this paper, combined with other techniques
for longer tracks \cite{Gunion:1999jr}.
The LEP-II doublet search proposed by Thomas and Wells
\cite{Thomas:1998wy} would be sensitive to 70 GeV masses, with
splittings on the order of $300$~MeV. As a linear collider is free of large low-energy
hadronic background from the underlying event, these authors proposed a search for
soft pions from the charged particle's displaced vertex. As with the Tevatron proposal,
this search has also not been performed.

Given the weakness of current bounds, any improvement would be
worthwhile. Surprisingly, the only significant bound improvements
are likely to be on triplet states although with enough luminosity
some improvement on bounds for other multiplets might be possible.
We will show that we might extend the mass reach for  vector-like
triplets to 500 GeV under optimistic high-luminosity LHC scenarios,
and that even in early running the bounds can be significantly
improved.

We now briefly outline the experimental issues and our proposed
strategy. Due to the relatively rapid decay (compared to the length
scales of the detector) of the charged states ($X^\pm$) to the
invisible lighter neutral state ($X^0$) plus very low energy pions or
electrons, experimental detection is extremely difficult. When rapid
decay to pions dominates ($\Delta M
>m_\pi$), pair production of $X^ \pm X^\mp + X^\pm X^0$ is
effectively invisible, since there is not a large amount of energy
deposited in any calorimeter layer or the muon system. Due to the
lack of energy deposition, such events are not triggered upon and so
would not even be written to tape. Smaller splittings ($\Delta M <
m_\pi$) would be less troublesome, as the longer lifetime resulting
from phase space suppression to $X^0e^\pm \nu_e$ states allows the
$X^\pm$ to pass through the calorimeter and even the muon system,
appearing as a slow, heavy, massive charged particle as it does so
\cite{Gunion:1999jr}. We will focus on the more difficult short
track search in this paper.

When the splitting is large enough to permit  the pion decay mode,
associated production of either photons or jets at high $p_T$  will
be needed to pass the detector trigger. Comparing the two processes
 at the LHC,  only jets would yield sufficient rate. One can trigger
on either high $p_T$ jets or on missing transverse energy (MET) (or
on both), where the missing $E_T$ comes from a pair of $X$ particles
recoiling against one or more jets. Once the events are triggered on
and recorded, off-line analysis will be used to look for the short
charged state tracks (prior to decay).

If the mass splitting is close to the pion mass, the typical $c\tau$
of the charged particles is on the order of $10~$cm. The central
trackers of both CMS and ATLAS consist of multiple detector layers spaced
$\sim 5$~cm to $\sim 50$~cm from the beam pipe. For this range of
mass parameters, the charged track may be reconstructed off-line by
requiring $\geq 3$ hits in the silicon tracker, with the main
background presumably being the combinatoric one from hits that
accidentally line up. Specialized reconstruction techniques would be
required since current algorithms require energy deposition in the
calorimeters or muon system to confirm a track. Using both
charged-charged and charged-neutral production and studying relative
rates could help reduce the background.

For larger splittings, $200~\mbox{MeV} \lesssim \Delta M \lesssim 1~\mbox{GeV}$,
 the decay length is $\lesssim 1~$cm so the tracks would be
too short to reliably affect even one layer of the detector. While
the pions in this case may have high enough transverse energy
($E_T$)
 to be efficiently detected, the huge hadronic background present in
 every bunch-crossing at the LHC completely dwarfs any possible signal.
 Detection of such particles would  be possible only at lower masses (and
 thus higher production cross sections) as the number of long lifetime events
reaching three or more layers is exponentially suppressed.

In Section~\ref{sec:models} we outline the models and spectra used
in the remainder of the paper. We consider new $SU(2)_L$ triplets
and new $SU(2)_L$ doublets, the phenomenology of which is detailed
in Section~\ref{sec:triplet}. These simple models are highly
predictive and suggest important qualitative and quantitative
differences in the strategy for discovery at the LHC. Perturbations
on these simple models (including the case of anomaly mediated
winos) will be discussed in Section~\ref{sec:complications}.

\section{Models} \label{sec:models}

As stated in the introduction, in this paper we are not attempting
to create a full model of TeV-scale physics that addresses the known
issues with the Standard Model: electroweak symmetry breaking
(EWSB), dark matter, naturalness and the hierarchy problem. Rather,
we take a minimalist approach, introducing vector fermions charged
under $SU(2)_L$ (with appropriate hypercharge to have a neutral
component) and investigating the reach of the LHC in discovering
such particles. It is interesting that what amounts to heavy
vector-representation leptons are so difficult to detect.

The Lagrangian of our minimal model consists of just the standard
kinetic and mass terms:
\begin{equation}
{\cal L} = i \bar{X} \slashed{D} X - M \bar{X}X. \label{eq:lagrangian1}
\end{equation}
Here the $X$ fields are vector fermions (that is, both left- and
right-handed components have the same quantum numbers) in a
multiplet $m=2,3,\ldots$ of $SU(2)_L$. We are interested in the
cases where one of the components of the multiplet has $Q_{\rm
EM}=0$ after electroweak symmetry breaking. This limits the choices
of hypercharge $Y$; for $m=2$, $Y=\pm 1/2$, $m=3$, $Y=0,\pm 1$; {\it
etc}. For specificity and because they are sufficient to illustrate
qualitatively different detector phenomenology, we choose to
consider two cases: $m=2$, $Y=1/2$ doublets $X_2$, and $m=3$, $Y=0$
triplets $X_3$. We will, however, comment briefly on the
phenomenology of other possibilities in this section.

For doublets there are four degrees of freedom:
\begin{equation}
X_2 = \left(\begin{array}{c} X_2^+ \\ X_2^0 \end{array}\right) \ ,
\
\bar{X}_2 = \left(\begin{array}{c} \bar{X}_2^0 \\ X_2^-
\end{array}\right). \label{eq:2dof}
\end{equation}
We choose triplets in a real representation of $SU(2)_L$. This
requires only three degrees of freedom:
\begin{equation}
X_3 = \left(\begin{array}{c} X^+_3 \\ X^0_3 \\ X^-_3 \end{array}\right). \label{eq:3dof}
\end{equation}
Notice that we could equally well have chosen six degrees of freedom for the triplets
(by introducing an $\bar{X}_3$). However, later we will draw connections between the triplet
model and anomaly-mediated winos, so we choose to focus on the Majorana case.
In all cases, the particle representations are vector-like and so the new states do not contribute to anomalies.

\begin{figure}[ht]
\includegraphics[width=0.5\columnwidth]{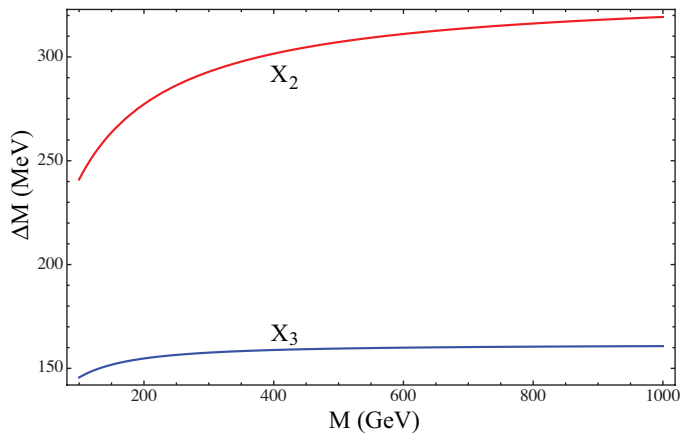}
\caption{Splittings between the charged $X^\pm$ and neutral $X^0$ states for $SU(2)_L$
doublets $X_2$ (red line) and triplets $X_3$ (blue) as a function of neutral state mass $M$. \label{fig:deltam}}
\end{figure}

With the Lagrangian of Eq.~(\ref{eq:lagrangian1}), all components of
the $X$ multiplet are degenerate at tree level. Splittings between
the charged and neutral components arise from loops of the $W^3$
gauge boson (or $Z/\gamma$ loops after EWSB). The mass splitting
$\Delta M$ between the state with charge $Q$ and the neutral state
is \cite{Cirelli:2005uq}
\begin{equation}
\Delta M = \frac{\alpha M}{4\pi} \left\{ Q^2 f\left(\frac{M_Z}{M}\right)+\frac{Q(Q-2Y)}
{\sin^2\theta_W}\left[f\left(\frac{M_W}{M}\right)-f\left(\frac{M_Z}{M}\right) \right]\right\}, \label{eq:deltam}
\end{equation}
where the function $f$ is
\begin{equation}
f(x) = \frac{x}{2}\left[ 2x^3 \ln x-2x+\sqrt{x^2-4} (x^2+2)\ln (x^2-2-x\sqrt{x^2-4})/2 \right] \label{eq:fdeltam}.
\end{equation}
The resulting mass splittings $\Delta M(M)$ between the charged and
neutral states for the doublets $X_2$ and triplets $X_3$ are displayed in Fig.~\ref{fig:deltam}.

These mass splittings have several features of interest. As
expected, the scale of $\Delta M$ is down from $M$ by a loop factor
$\alpha/4\pi$, leading to ${\cal O}(100~\mbox{MeV})$ splittings from
TeV-scale masses. This has important consequences for the decay
width of $X^\pm \to X^0 +(\mbox{SM fields})$.

The electron mode dominates for $\Delta M < m_\pi$, but once
kinematically allowed, the single pion mode dominates
 until $\Delta M \gtrsim 1$~GeV \cite{Chen:1995yu}. Beyond
this splitting, the $\pi^\pm \pi^0$ and other hadronic modes become
important. As can be seen from Fig.~\ref{fig:deltam}, the scale of
$\Delta M$ is about $300$~MeV for doublets and approximately
$150$~MeV for triplets allowing us to ignore all modes except the
single pion (Eq.~(\ref{eq:piwidth})). Since $\Delta M \sim
\Lambda_{\rm QCD}$, only the two-body channel $X^\pm \to X^0
\pi^\pm$ is open, and two-body kinematics causes this mode to
dominate for this range of mass splittings.

A perhaps counter-intuitive result of Eq.~(\ref{eq:deltam}) is that
the splitting of the triplets is smaller than that of the doublet.
This is because of the non-zero hypercharge required for a neutral
component of the $X_2$ multiplet. Since the splitting is
smaller for $X_3$ and is only marginally larger than
$m_\pi$, the triplet lifetime is much longer than that of $X_2$.

From Eq.~(\ref{eq:deltam}), we find a generic result for vector
fermions in a representation $m$ of $SU(2)_L$ with a neutral ground
state. The $X^\pm-X^0$ splittings for even $m$ will be ${\cal
O}(300~\mbox{MeV})$ for $M \gtrsim 200$~GeV, while for odd $m$,
$\Delta M$ will be ${\cal O}(150~\mbox{MeV})$, assuming $Y=0$. Of
course, when $Y\neq 0$, the odd $m$ splittings will generally be
larger than that of the even representations.

The $X^\pm$ decay widths into various final states have been
calculated in several sources for doublets \cite{Thomas:1998wy},
triplets \cite{Chen:1995yu,Ibe:2006de}, and higher multiplets
\cite{Cirelli:2005uq}. The most relevant for our purposes are the
decays to a single pion or $\ell \nu_\ell$ pair, which have widths
of
\begin{eqnarray}
\Gamma(X^\pm \to X^0 \pi^\pm) & = & \frac{A G_F^2}{4\pi} \cos^2 \theta_c f_\pi^2
\Delta M^3 \sqrt{1-\frac{m_\pi^2}{\Delta M^2} }\label{eq:piwidth} \\
\Gamma(X^\pm \to X^0 \ell^\pm \nu_\ell) & = & \frac{A G_F^2}{60
\pi^3} \Delta M^5
\sqrt{1-\frac{m_\ell^2}{\Delta M^2} }
P\left(\frac{m_\ell}{\Delta M} \right)\label{eq:leptonwidth},
\end{eqnarray}
where $f_\pi \approx 130$~MeV is the pion decay constant, $\theta_c$ is the Cabibbo angle,
$A=m^2$ for even representations $m$ of $SU(2)_L$, for odd representations $A=4(m^2-1)$, and the function $P(x)$ is defined
as
\begin{equation}
P(x) = 1 - \frac{9}{2} x^2 - 4 x^4 + \frac{15 x^4}{2\sqrt{1-x^2}} \tanh^{-1} \sqrt{1-x^2} \label{eq:P}.
\end{equation}

\begin{figure}[ht]

\includegraphics[width=0.4\columnwidth]{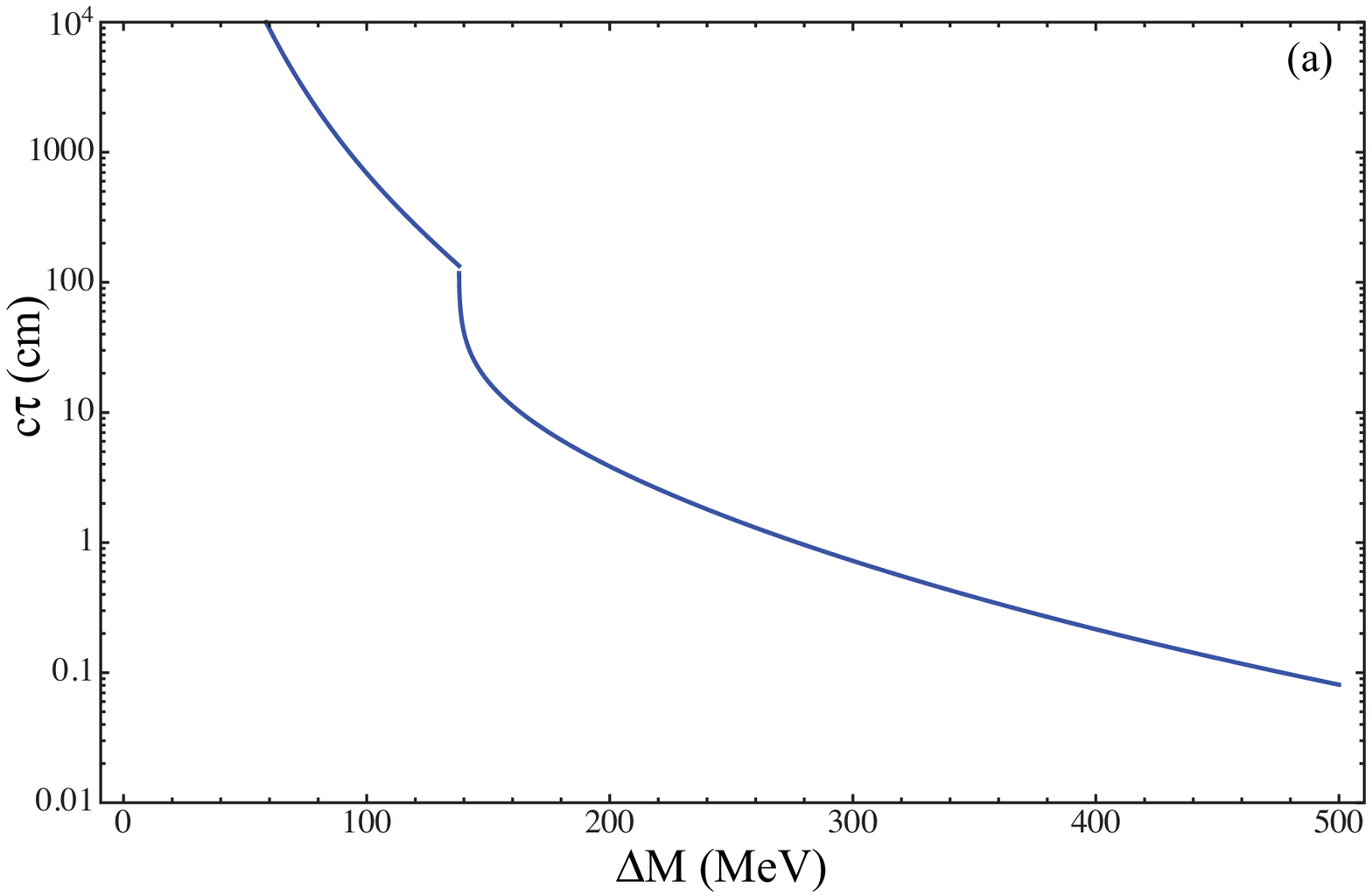}\includegraphics[width=0.4\columnwidth]{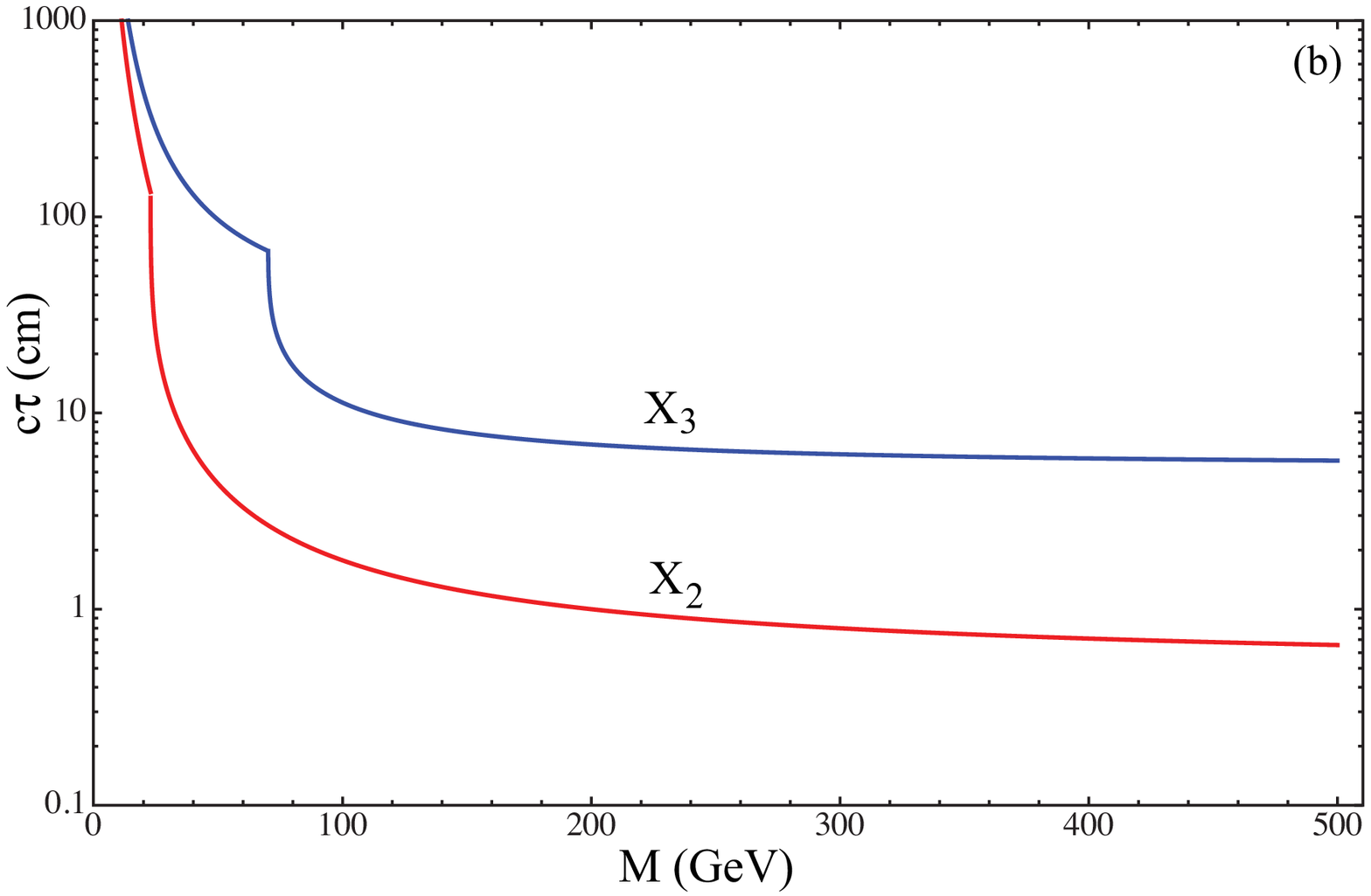}

\caption{Left: Lifetime of generic $X^\pm$ decaying to the final states $X^0+\pi^\pm$
or $X^0+\ell^\pm+\nu_\ell$ as a function of $X^\pm-X^0$ mass splitting $\Delta M$.
Here $\ell = e/\mu$.  We have assumed that the $X$ couples to $W^\pm$ with 
the same strength as a triplet  ({\it i.e.}~$A=8$ in Eqs.~(\ref{eq:piwidth}) and 
(\ref{eq:leptonwidth}) The `kink' structure at $\Delta M = m_\pi$ is caused by the pion
channel opening. Right: Lifetime of doublets $X^\pm_2$ and triplets $X^\pm_3$ decaying
to the neutral states $X^0_2/X_3^0$ as a function of the neutral state mass $M$,
assuming the mass splittings of Eq.~(\ref{eq:deltam}).  \label{fig:ctau}}
\end{figure}

The resulting lifetime $c\tau$ for a  multiplet with splitting
$\Delta M$ is shown in Fig.~\ref{fig:ctau}(a), while the lifetime of
the doublets and triplets as a function of the $X^0$ mass $M$ is
shown in Fig.~\ref{fig:ctau}(b). As expected, there is a noticeable
`kink' structure when the pion channel opens at $\Delta M = m_\pi$.
For $X_2^0(X_3^0)$ masses above $200$~GeV, the lifetime is
remarkably constant at $\sim 1(10)~$cm.

When viewing $c \tau$ keep in mind that this is not the physical
distance that will be measured in the detector. When the factors
$\beta \gamma$ are included, the path lengths generally increase by
a factor of a few. However, since only the transverse distance is
relevant to track determination, the physically relevant distance
will be comparable to the $c \tau$ values presented below. These
numbers will be made more precise when we consider actual event
simulations.

The simple Lagrangian of Eq.~(\ref{eq:lagrangian1})
provides a very predictive model for experimental searches. Charged
particles will be created (either in pairs or in conjunction with a
neutral state), travel ${\cal O} (10~\mbox{cm})$ for triplets or
${\cal O}(1~\mbox{cm})$ for doublets, and decay into an invisible
$X^0$ and a very low energy pion.
A new vector triplet with mass of ${\cal O}(\mbox{TeV})$ would
have a mass splitting between the $X^+_3$ and $X^0_3$ states that is
only slightly above the pion mass. From the point of view of an LHC
experiment (ATLAS or CMS), the $X^+_3$ (in principle visible as an
ionizing charged track in the central tracker), will decay
into an invisible $X_3^0$ and a pion after ${\cal
O}(10~\mbox{cm})$.

In the rest frame of the decaying particle, the energy of this
outgoing pion is
\begin{equation}
E_\pi =  \frac{ 2M\Delta M+\Delta M^2+m_\pi^2}{2(M+\Delta M)} \sim
\Delta M \sim 150~\mbox{MeV}. \label{eq:epion}
\end{equation}
The experimentally relevant parameter is not $E_\pi$, but the
transverse pion energy $E_{T}$, which is of course smaller. While we
may expect some gain in available energy due to relativistic boosts
of the parent particle, this is not significant for the range of $M$
we are interested in. For these ranges of $E_T$, the pions will
typically not even register in the detector. Therefore the search
strategy in this case will be to look for stubs from charged tracks
after triggering on missing or jet energy.

Doublets on the other hand generally have larger splittings. This
implies the lifetime and track stub length is expected to be shorter
than that of triplets, making the
 search for stubs even more challenging. The pion
transverse momentum is generally bigger, in the range of efficient detection
at ATLAS and CMS \cite{PausTalk}. However, even though the pions from these decays
will be isolated from jets and have large impact parameters, the extremely
large background of pions from the underlying event (and pile-up
during high luminosity runs) will almost certainly make such detection
impossible.

For larger splittings on the order of 1 GeV, the decay mode includes
multiple pions and other hadronic states. Jets with $E_T$ on the
order of a GeV and large impact parameters are a key part of $b$-jet
tagging \cite{Bayatian:2006zz}. Thus, once $\Delta M \sim {\cal
O}(1~\mbox{GeV})$, standard triggers should be capable of
registering their presence. Models with such large splittings are
not considered in this paper. However, we note that this nonetheless
leaves a range of $\Delta M$ from $\sim 200$~MeV to 1 GeV where
detection of new multiplets is extremely unlikely.

Although generically the relation Eq.~(\ref{eq:deltam})
holds, there are important exceptions.
In Section~\ref{sec:complications} we will also consider an
interpolating scenario between that of the doublets and the triplets
where $M$ and $\Delta M$ are less rigidly connected than in
Eq.~(\ref{eq:deltam}). This is important since the key parameter for
all these searches is the splitting $\Delta M$, which controls the
branching ratios, lifetimes, and energies of the final states in the
detectors.

As we shall see, the charged particle width $\Gamma$  essentially
determines the detectability of a model. This parameter is set by a
combination of the splitting and the couplings to $W^\pm$ (see
Eq.~(\ref{eq:piwidth})). While the splitting for higher odd
multiplets remains small (for zero hypercharge), the larger
couplings increase $\Gamma$ and thus make those scenarios more
difficult to discover, though we would be aided by the larger
production cross section.

As an explicit example, consider an $m=5$ multiplet with zero
hypercharge. This consists of neutral, $\pm e$ and $\pm 2 e$ charged
states. The splitting between the neutral and singly charged states
is the same as in the $m=3$ triplet, while the doubly charged states
are $\sim 650$~MeV heavier than the $X_5^0$.
These doubly charged $X_5^{\pm\pm}$ states would rapidly decay first
to $X_5^\pm$ and then to the neutral state. Due to smallness of the
mass splitting and resulting low energy of the SM decay products,
it is unlikely that
the byproducts of such a cascade would be visible at the LHC, or
that the heavy charged particle would be deflected by a noticeable
amount. This makes the $X^{\pm\pm}_5$ production act effectively as
an increase in the $X^\pm_5$ production cross section. The $X_5^\pm$
then decays to $X_5^0\pi^\pm$ more quickly than an equal mass
$X_3^\pm$, due to the larger coupling of higher multiplets to
$W^\pm$. For a generic odd multiplet $m$, the width goes as $m^2-1$,
while for even $m$ widths are proportional to $m^2$.

As a result, an $m=5$ multiplet would decay faster than a triplet by
a factor of eight. The detector phenomenology would be similar to
that of the doublet but with a larger apparent production of singly
charged heavy states and slightly longer tracks.\footnote{A future
linear collider, free of the large hadronic background from the
underlying event, might be able to find the pions from cascade
decays of high multiplets. Such a search is beyond the scope of this
paper.} Surprisingly, triplets are therefore the easiest multiplet
to discover, as they have the longest decay length. Below we
consider both triplets and doublets, the latter of which can still have an
improved bound over current constraints, even with their larger mass
splitting.

\section{Signature of Triplets and Doublets} \label{sec:triplet}

As we've seen, the small pion $E_T$ is due to the splitting $\Delta
M$ being close to the pion mass. This coincidence, while eliminating
the final state particles as a detection path, leaves another
possibility open. Because of the constrained phase space available
for the decay products, the lifetime of the charged state is
relatively long. Therefore, searches for such particles could look
for a charged particle traveling through the central tracker and
then disappearing -- a stub in the tracker. With two charged final
state particles there would of course be two stubs.

Such searches will of course be challenging.  While a $c\tau$ of 10
cm is huge compared to the lifetimes of SM particles, it is  small
compared to the physical size of the detector. The ATLAS central
tracker consists of three silicon pixel layers in the barrel at a
distance of $5.05$, $8.85$, and $12.25$ cm from the beam, followed
by four layers of the silicon central tracker (SCT) at $29.1$,
$37.1$, $44.3$, and $51.4$ cm \cite{Aad:2009wy}. Thus, most of the
$X^+_3$ decays would occur before the SCT, and only $\sim e^{-5}$ of
all particles would reach the Transition Radiation Tracker (TRT) at
$56.3$~cm. Despite the great differences in design in the outer
sections (most significantly in the muon systems), the CMS detector
is similarly configured in the central tracking region, with three
pixel layers at $4.4$, $7.3$, and $10.2$~cm, while the first layer
of the silicon strip tracker is at $22$~cm \cite{Bayatian:2006zz}.
As the two central tracking regions are so
similar, we use the slightly more conservative (for our purposes) ATLAS distances.

Since the only visible signature of $X^+X^-$ production is stubs,
production of these particles would be missed given current triggers.
In order to record production events, we therefore require at least one associated jet
produced by initial state radiation, {\it i.e.}~$pp\to X^+_3X^-_3
+\mbox{jets}$. The idea would be to trigger on jet energy or missing
energy, but identify the stubs off-line to distinguish signal events
from the abundant jet or missing energy background.

As production proceeds via weak couplings, the additional
requirement of associated jets greatly reduces an already small
cross section (especially for the larger values of $M$). We
therefore consider only  unprescaled triggers. We choose a missing
transverse energy (MET) trigger, requiring MET$>65$~GeV in the high
luminosity mode \cite{Adam:2005zf}. We select MET rather than jet
$p_T$ because of the lower (net) jet energy specified to pass the
trigger cut.

Nonetheless,  this trigger may turn out to be overly aggressive, so
the scaling of production cross section with a variety of trigger
options for  $M = 300$~GeV is shown in Table~\ref{tab:triggersigma}.
Also in Table~\ref{tab:triggersigma}, we give the cross sections for center of mass
energies of 10 TeV and 13 TeV, for the various trigger selections.

\begin{table}[ht]

\begin{tabular}{|c|c|c|c|c|}
\hline
$\sqrt{s}$ (TeV) & MET$> 65$~GeV & MET$>80$~GeV & Jet $p_T > 110$~GeV & jet $p_T>150$~GeV \\ \hline \hline
13 & $50.2$~fb & $44.1$~fb & $34.2$~fb & $23.9$~fb \\ \hline
10 & $26.5$~fb & $23.1$~fb & $17.1$~fb & $11.5$~fb \\ \hline
\end{tabular}

\caption{The production cross section of $X_3\bar{X}_3$ pairs at
$\sqrt{s} = 13$~TeV and $10$~TeV with $M=300$~GeV using four
different trigger menus. These are triggers on MET$>65$~GeV, MET$>
80$~GeV, highest jet $p_T > 110$~GeV, and a highest jet $p_T >
150$~GeV. All cross sections were calculated using MadGraph
\cite{Stelzer:1994ta} and Pythia \cite{Sjostrand:2006za}. See text for further details.
\label{tab:triggersigma}}
\end{table}

We see that we expect a reasonable number of such events for particle
masses of order a few hundred GeV.  We now consider the cross
sections with the  restrictions on the stubs necessary to
detect the signal events.

The principal requirement is that both tracks must pass through at
least three layers of the central tracker
so that the stubs are sufficiently long to be found in an off-line
analysis. This requires the transverse physical track lengths
($\beta \gamma c \tau \sin \theta$, where $\theta$ is the polar
angle measured from the beam line) to be greater than 12 cm. In the
remainder of this paper, we will refer to this quantity as
transverse $c \tau$, or $Tc \tau$.
One should bear in mind that
conventional track reconstruction algorithms require hits in all
the central tracker layers. Without a detailed detector simulation or data
analysis it is difficult to know with certainty that three hits will
be sufficient to identify signal events and distinguish them from
background, but in the interest of maximizing signal we first
explore this possibility.

However, experimental groups might find more hits necessary for purposes of background
rejection, track reconstruction, and charge assignment (as longer
tracks provide a better handle on the track curvature, and thus sign of the charge).
The next layer (the fourth from the
center and the first of the SCT) lies at 30 cm at ATLAS. Due to the slightly more
compact design, this transverse distance would cross two additional
tracker layers at CMS, for a total of five hits. In both cases
we expect a serious decrease in the cross section of events with 4 hits on both tracks.
We will show the expected cross section requiring $Tc\tau>30$~cm on
both tracks, as well as events with one stub of $12$~cm and
30~cm on the other. Thus, in addition to the $\mbox{min}[Tc\tau]>12$~cm
cut, we also show results for $\mbox{min}[Tc\tau]>30$~cm and $\{\mbox{min}[Tc\tau]>12~\mbox{cm},
\mbox{max}[Tc\tau]>30~\mbox{cm}\}$.

In all cases, we  require both tracks to have $|\eta|<2.5$ so that
they sit in the central barrel, where the tracking layers are
closest to the interaction point. Our final requirement is that
$\Delta R>0.4$, where $\Delta R$  is the minimum $\phi-\eta$ angular
distance  between both tracks and any jet, so that the two tracks
are isolated from all jets in the event.

\begin{figure}[t]
\includegraphics[width=0.5\columnwidth]{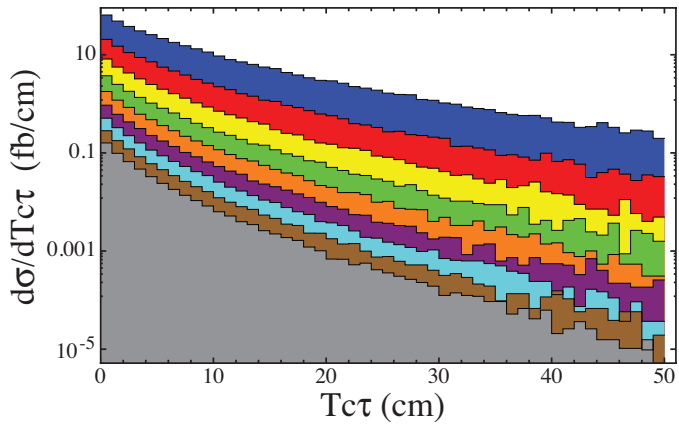}

\caption{Differential cross section versus transverse path length
$Tc\tau$ of $X^\pm_3$ from $X^+_3X^-_3+\rm{jets}$ events at $\sqrt{s}=13$~TeV,
triggering on $65$~GeV MET signature. Cuts of $\Delta R >0.4$ and $|\eta|<2.5$
have been applied, but no cut on $Tc\tau$ has been added (see text for more detail).
Color coding denotes mass of $X_3^0$ state: blue is 200~GeV, red is 300~GeV,
yellow 400~GeV, green 500~GeV, orange 600~GeV, purple 700~GeV, cyan
800~GeV, brown 900~GeV, and grey 1000~GeV. \label{fig:tripletctau}}
\end{figure}

\begin{figure}[t]
\includegraphics[width=0.5\columnwidth]{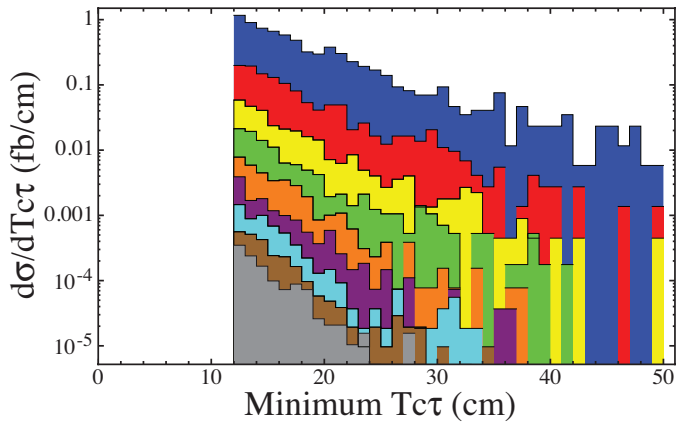}

\caption{Differential cross section versus the minimum value of $Tc\tau$ of $X^+_3X^-_3$
pairs. $X^+_3X^-_3+\rm{jets}$ events were generated at $\sqrt{s}=13$~TeV, with
65~GeV MET triggering. Cuts on $\Delta R > 0.4$, $|\eta|<2.5$, and $\mbox{min}[Tc\tau] >
12$~cm have been applied. Color coding for $X_3$ mass as in Fig.~\ref{fig:tripletctau}.
 \label{fig:tripletctaulong}}
\end{figure}

In Figure~\ref{fig:tripletctau}, we plot the differential cross section for $X^+ X^- +
{\rm jets}$  production as a function of $Tc\tau$ for each event
for $M=200$, $300$, $\ldots$, $1000$ GeV to give an idea of the distribution of
path lengths.
In Figure~\ref{fig:tripletctaulong}, for the same range of masses, we plot the
differential cross section as a function of
minimum $Tc\tau$ of each $X^+_3X^-_3$ pair for those
events passing the cuts above.

In Table~\ref{tab:tripletctausigma}, we give the integrated cross
sections for the events passing our cuts  as a function of mass.
We provide integrated cross sections for the $\mbox{min}[Tc\tau]>12$~cm,
 $\mbox{min}[Tc\tau]>30$~cm, and $\{\mbox{min}[Tc\tau]>12~\mbox{cm},
 \mbox{max}[Tc\tau]>30~\mbox{cm}\}$ cuts. Extrapolating these results, and assuming that five of these
isolated, opposite sign, three-hit tracks are sufficient for
discovery, an integrated luminosity of $10$~fb$^{-1}$ is sufficient to discover $X_3$
triplets with a mass of $\sim 350$~GeV at the LHC,
With $100$~fb$^{-1}$ triplets with masses up to $550$~GeV are
in reach. Requiring two tracks with four hits each ($\mbox{min}[Tc\tau]>30$~cm)
reduces our reach to 200~GeV for 10 fb$^{-1}$ and 300~GeV for 100~fb$^{-1}$.
In both cases, these results constitute significant advances on the current
exclusion regions.

\begin{table}[ht]
\begin{tabular}{|c|c|c|c|}
\hline
$X_3^0$ Mass & $\sigma_{12}$ & $\sigma_{30}$  & $\sigma_{12/30}$ \\
(GeV) & (fb) & (fb) & (fb) \\ \hline \hline
200 & 8.8 & $0.69$ &  $4.0$ \\ \hline
300 & 1.4 & $0.062$ & $0.45$ \\ \hline
400 & 0.32 & $0.013$ & $0.10$ \\ \hline
500 & 0.11 & $0.0037$ & $0.029$ \\ \hline
600 & 0.036 & 0.0005 & 0.0062\\ \hline
700 & 0.013 & 0.0001 & 0.0025 \\ \hline
800 & 0.0059 &  0.0000 & 0.0010 \\ \hline
900 & 0.0025 & 0.0000 & 0.0005 \\ \hline
1000 & 0.0012 & 0.0000 & 0.0002 \\ \hline
\end{tabular}

\caption{The integrated cross sections for $X^+_3X^-_3+\rm{jets}$ events in which
 $\mbox{min}[Tc\tau]>12$~cm ($\sigma_{12}$), as well as $\mbox{min}[Tc\tau]>30$~cm
 ($\sigma_{30}$); and $\mbox{min}[Tc\tau]>12$~cm, $\mbox{max}[Tc\tau]>30$~cm ($\sigma_{12/30}$),
 as a function of mass. Events were
simulated at $\sqrt{s}=13$~TeV with a MET trigger of $65$~GeV. In addition to the $Tc\tau$
requirement, events passed an $|\eta|<2.5$ cut on $X^\pm_3$ and an isolation cut of
$\Delta R >0.4$. The corresponding differential cross sections are shown in Fig.~\ref{fig:tripletctaulong}.
\label{tab:tripletctausigma}}
\end{table}

Figure~\ref{fig:tripletctaulong} and Table~\ref{tab:tripletctausigma} are the main results of
this section. At this point, we go into more detail on the techniques used to simulate the
production of the $X^+_3X^-_3$. Using our own model files, $X^+_3X^-_3/X^\pm_3X^0_3+\rm{jets}$
Feynman diagrams were generated using MadGraph \cite{Stelzer:1994ta}. At the partonic level,
we restricted ourselves to one or two jets (inclusive), due to computing constraints on
generating events with higher jet multiplicity.

For each choice of $X_3^0$ mass ({\it i.e.}~200, 300, $\ldots$, 1000
GeV), the mass splitting (and thus the mass of the $X_3^\pm$) and
decay width were calculated using Eqs.~(\ref{eq:deltam}) and
(\ref{eq:piwidth}). For each mass, $100,000$ events were generated
in MadGraph, requiring partonic jet $p_T > 15$~GeV, $|\eta|<5$, and
$H_T > 50$~GeV (here $H_T$ is defined as the magnitude of the vector
sum of all jet $p_T$). The large event sample was necessary in order
to gain meaningful statistics after the  severe cuts.

The underlying event and jet showering were provided by running the MadGraph samples
through stand-alone Pythia 6.409 \cite{Sjostrand:2006za}. Jet-finding was through Pythia's
built in cell algorithm, with an $R$ value of $0.4$. The $p_T$ cut-off for generation of the
underlying event was set at $15$~GeV. Decay of the charged triplets was done through
the standard Pythia decay routines, with the width and lifetime set by hand. MadGraph's
integrated Pythia and PGS options were not appropriate in this case, as PGS in particular
does not have the correct response to new stable particles which reach the detector.

\begin{figure}[t]

\includegraphics[width=0.5\columnwidth]{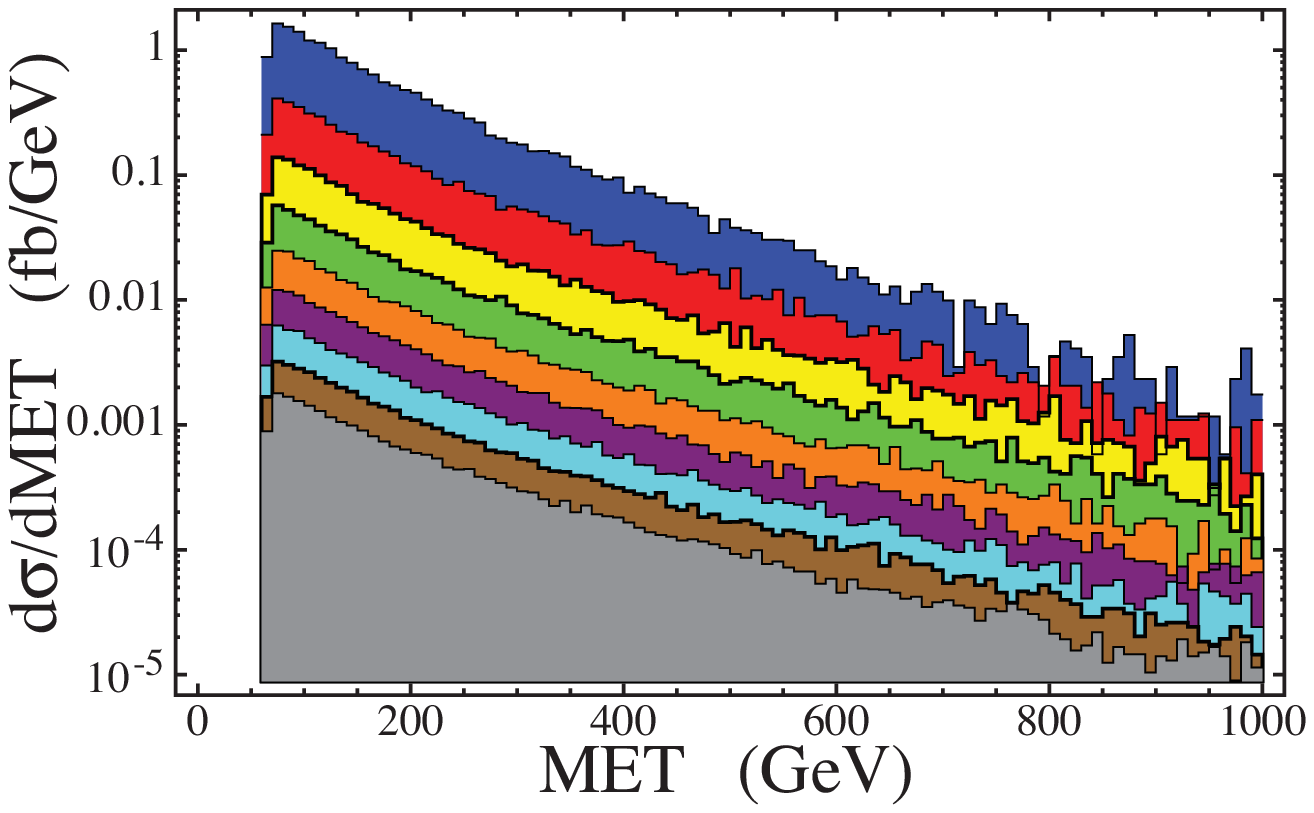}\includegraphics[width=0.5\columnwidth]{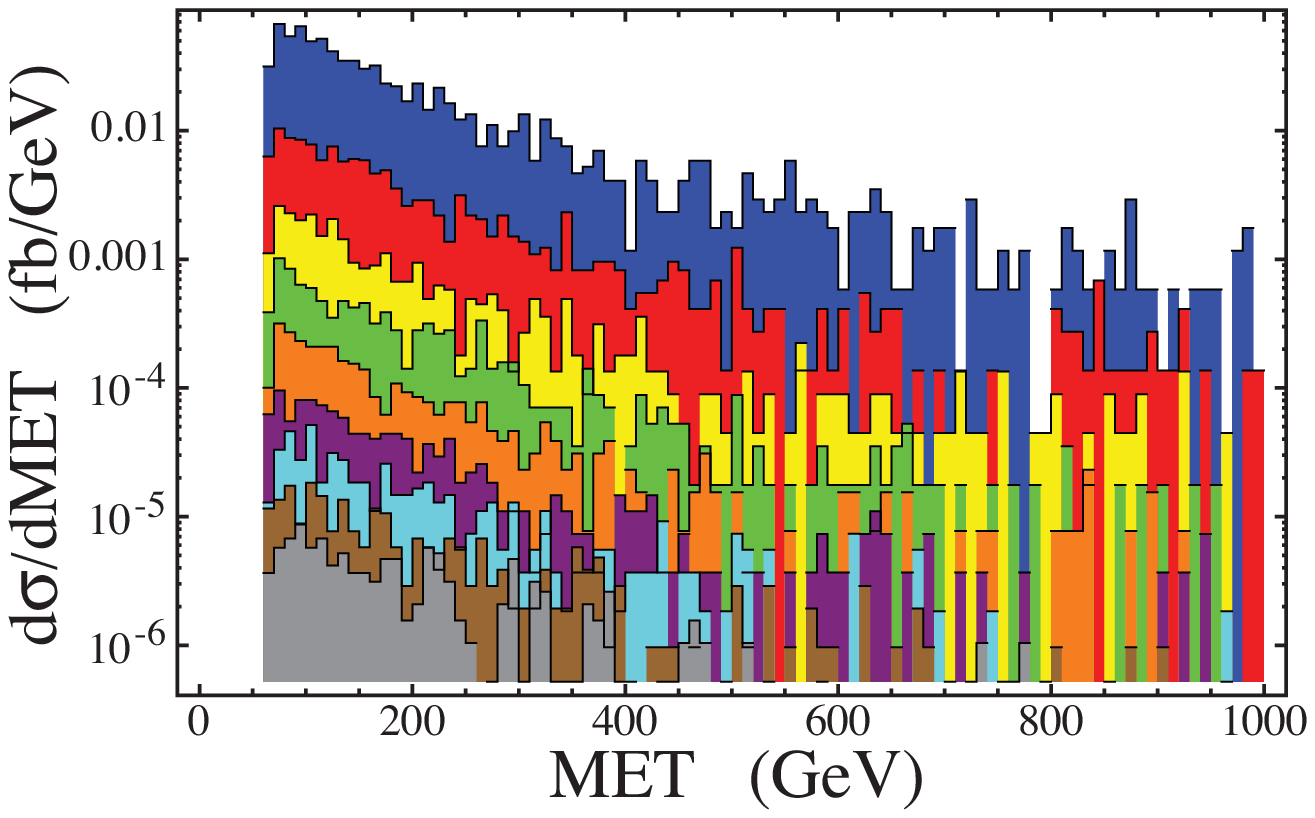}
\caption{Left: Missing transverse momentum (MET) of $X^+_3X^-_3+\rm{jets}$ events passing MET$>65$~GeV trigger as well as $X^\pm$ $|\eta|$ and isolation cut at $\sqrt{s} =13$~TeV. Right: MET distribution of those same events after application of a cut requiring both $X^+_3$ and $X^-_3$ have $Tc\tau>12$~cm. Color coding denotes choice of $X_3^0$ mass and is the same as in Fig.~\ref{fig:tripletctau}. \label{fig:tripletMET}}
\end{figure}

For all cuts based on jet $p_T$, the relevant values were
taken from the Pythia-identified jets, not from the parton-level.
The value of missing transverse energy for trigger purposes was
obtained for each event by taking the magnitude of the vector sum of
jet $p_T$ for all jets with $|\eta| <5$ and $p_T>20$~GeV.  After
determining whether an event passed the trigger, subsequent cuts on
$X^\pm_3$ isolation and $|\eta|$ were applied. Finally, a cut on the
minimum $Tc\tau$ of both charged triplets in $X^+_3X^-_3$ events was
performed. As outlined previously, the benchmark requirement of both
particles passing three layers corresponds to the minimum transverse
$c\tau$ being greater than
$12$~cm, while at ATLAS four hits on both tracks corresponds to 30~cm. A third option is to apply a cut
requiring one track to be longer than 12~cm, and the other longer than 30.
For our benchmark trigger, we use require MET$>65$~GeV.
For other trigger options, see Table~\ref{tab:triggersigma}. The differential distributions of MET
before and after the $Tc\tau$ cut are displayed in Fig.~\ref{fig:tripletMET}.

\begin{figure}[ht]
\includegraphics[width=0.5\columnwidth]{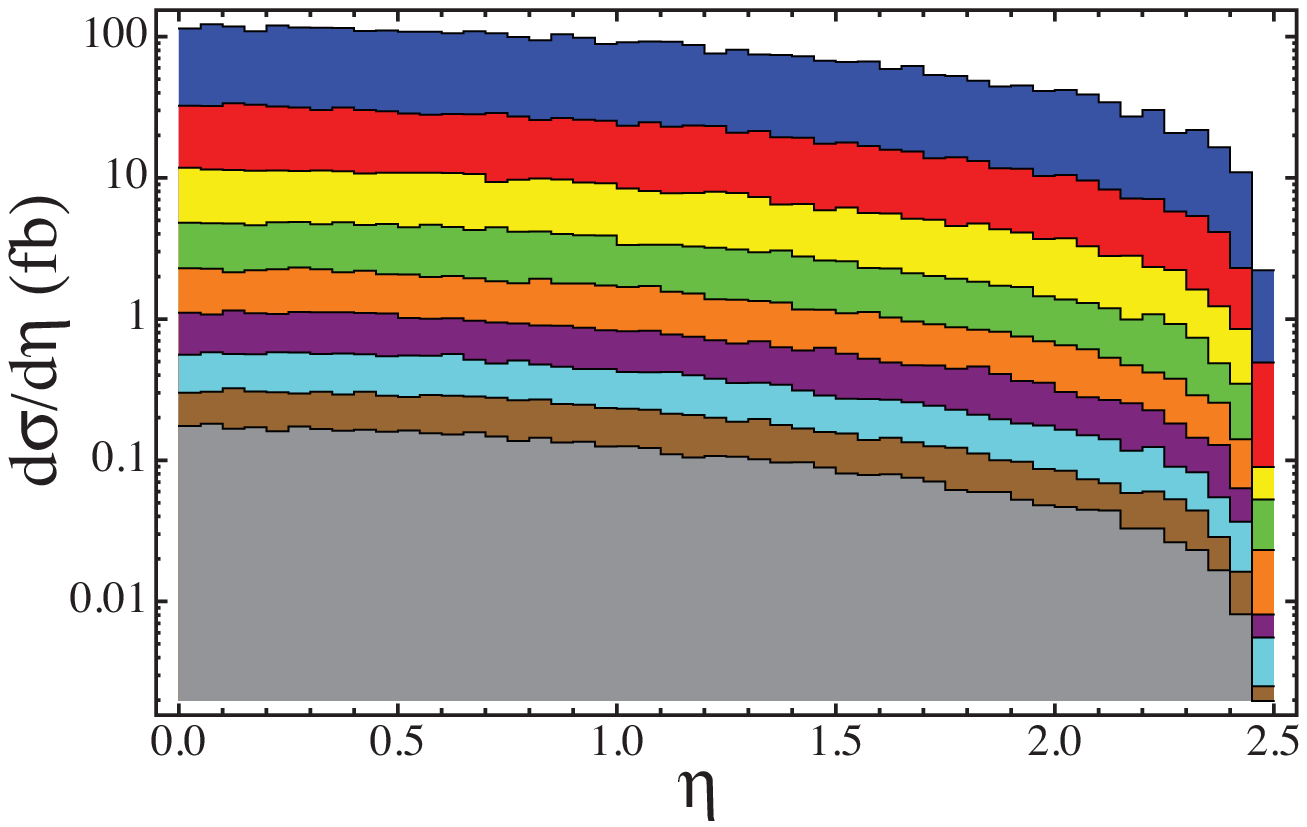}\includegraphics[width=0.5\columnwidth]{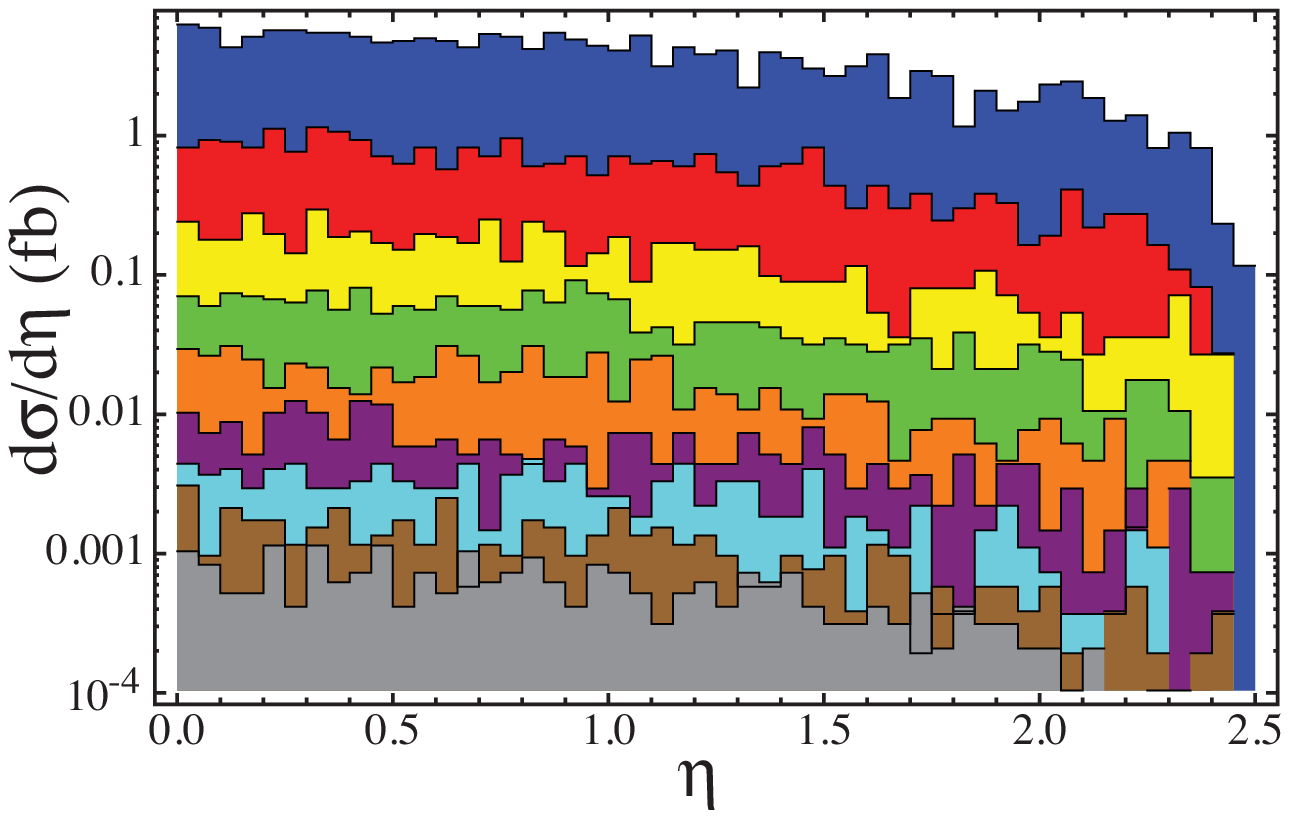}
\includegraphics[width=0.5\columnwidth]{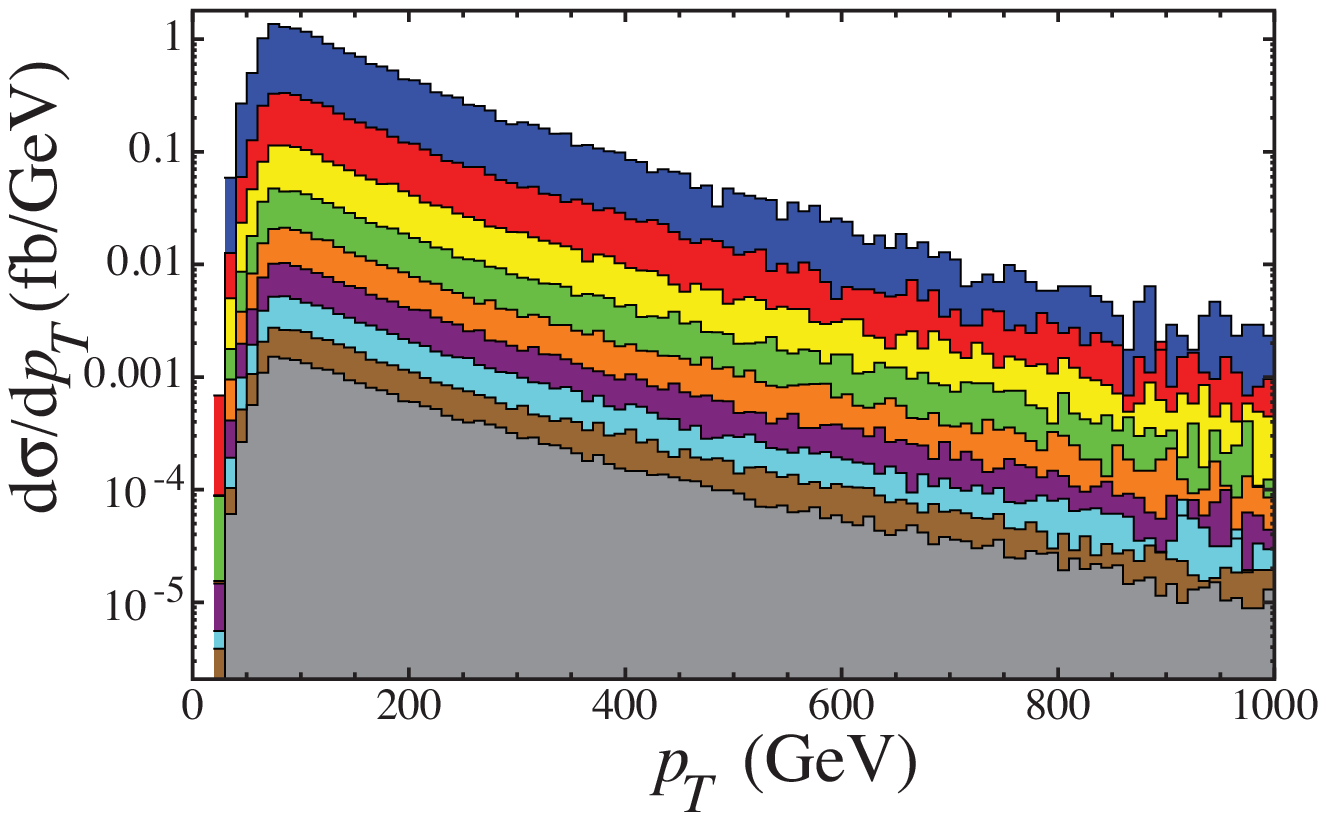}\includegraphics[width=0.5\columnwidth]{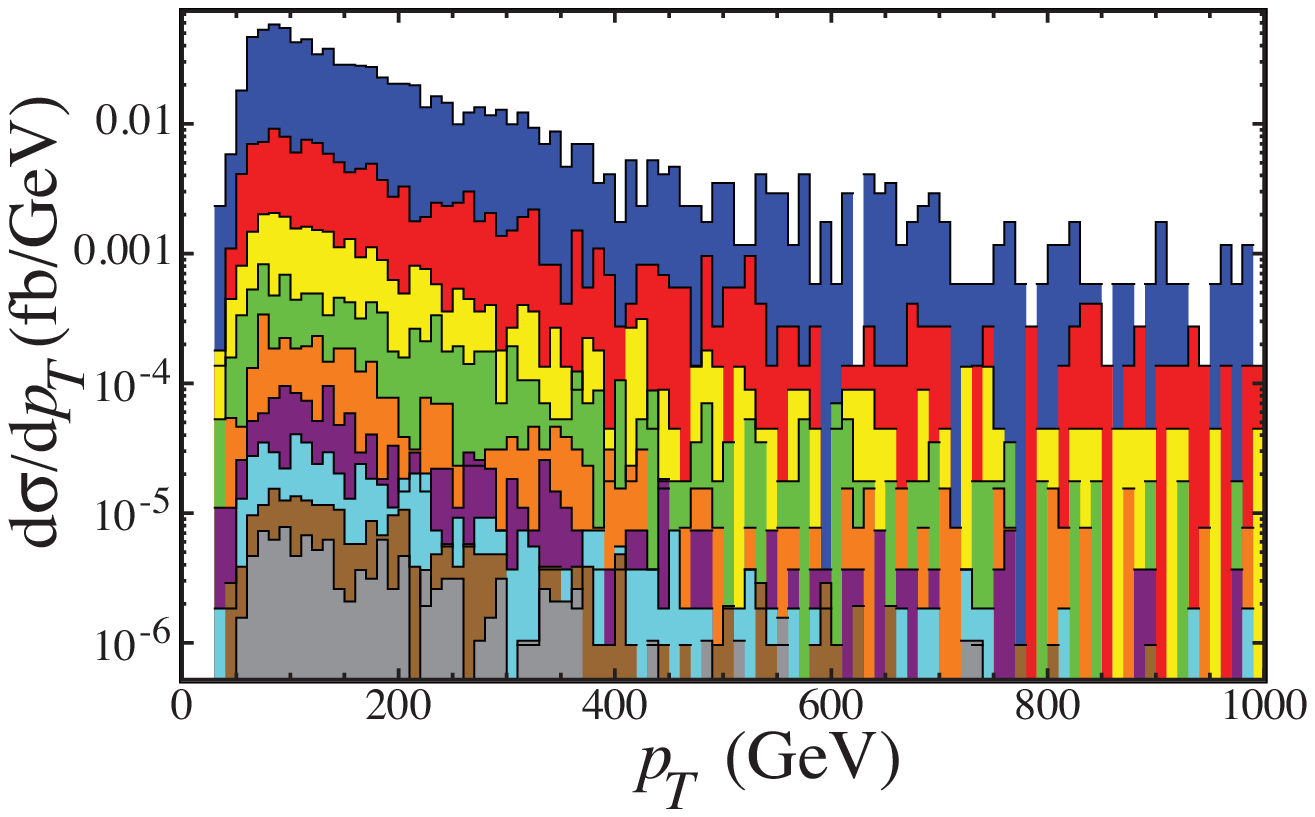}
\caption{Upper left: $|\eta|$ of highest $p_T$ jet for $X^+_3X^-_3 +\rm{jets}$ events passing MET$>65$~GeV trigger as well as $X^\pm$ $|\eta|$ and isolation cut at $\sqrt{s} =13$~TeV. Lower left: $p_T$ of highest $p_T$ jet from these events. Upper right: $|\eta|$ of highest $p_T$ jet from events that pass all cuts applied previous, plus the requirement that $Tc\tau >12$~cm for both $X^+_3$ and $X^-_3$. Lower right: $p_T$ of highest $p_T$ jet from those same events. Color coding denotes choice of $X_3^0$ mass and is the same as in Fig.~\ref{fig:tripletctau}. \label{fig:tripletjets}}
\end{figure}

\begin{figure}[ht]

\includegraphics[width=0.5\columnwidth]{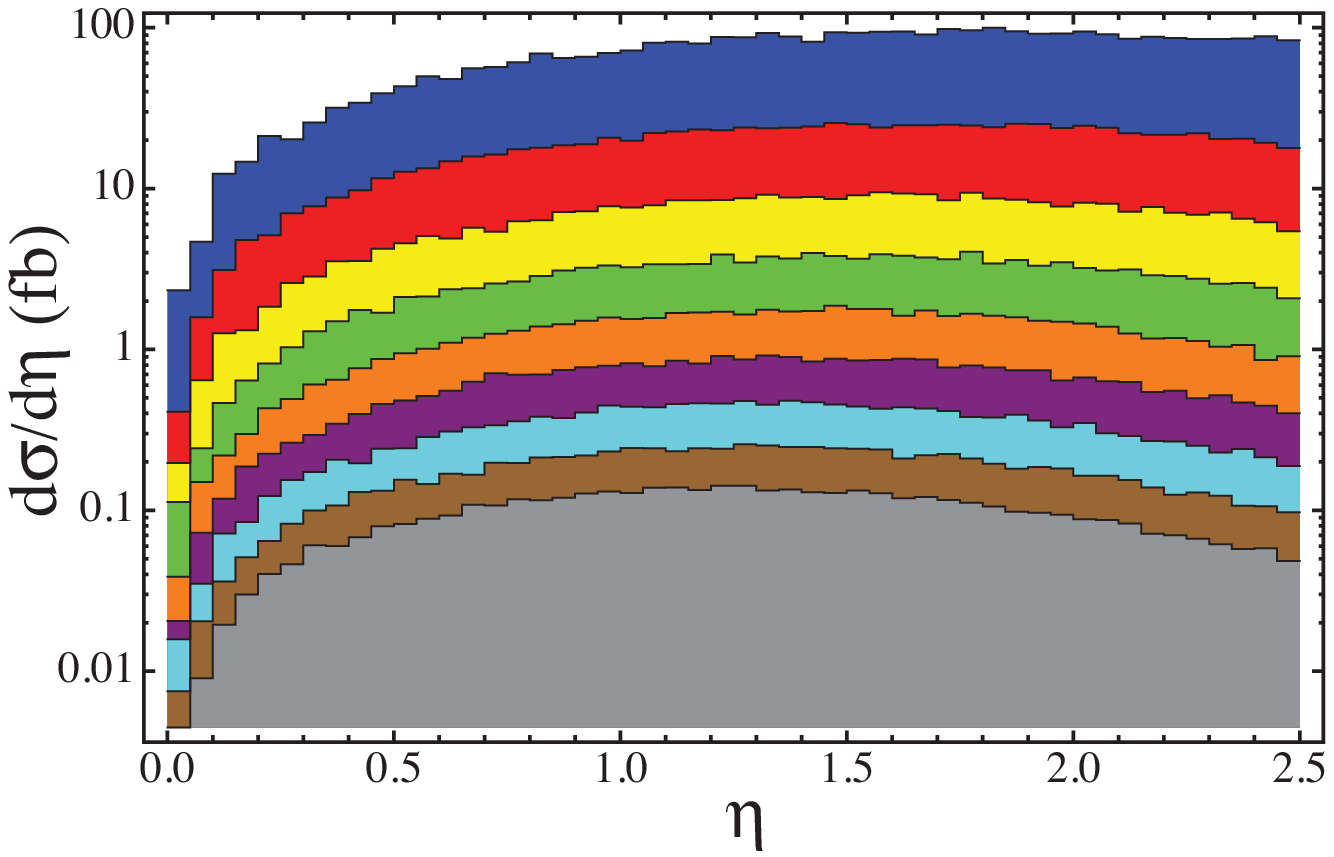}\includegraphics[width=0.5\columnwidth]{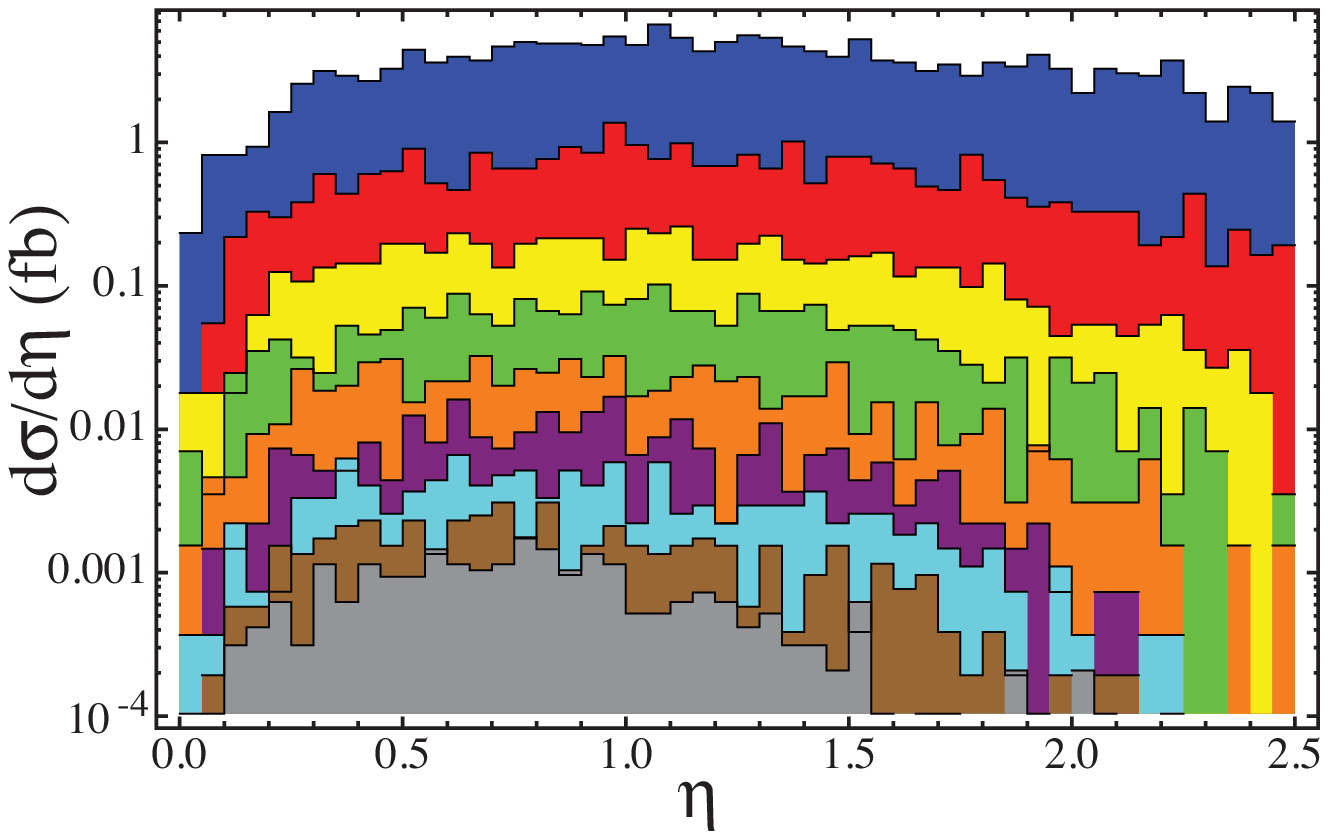}
\includegraphics[width=0.5\columnwidth]{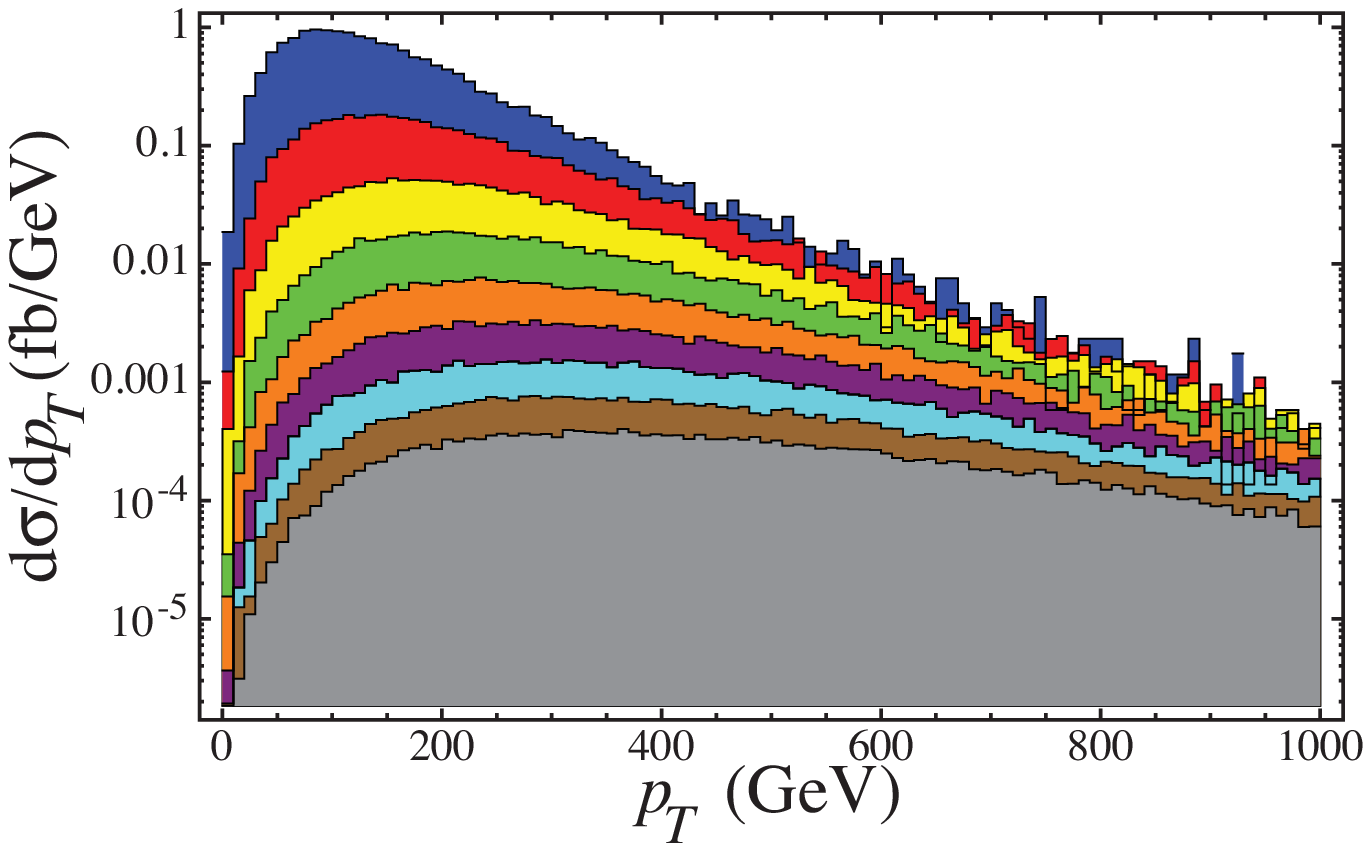}\includegraphics[width=0.5\columnwidth]{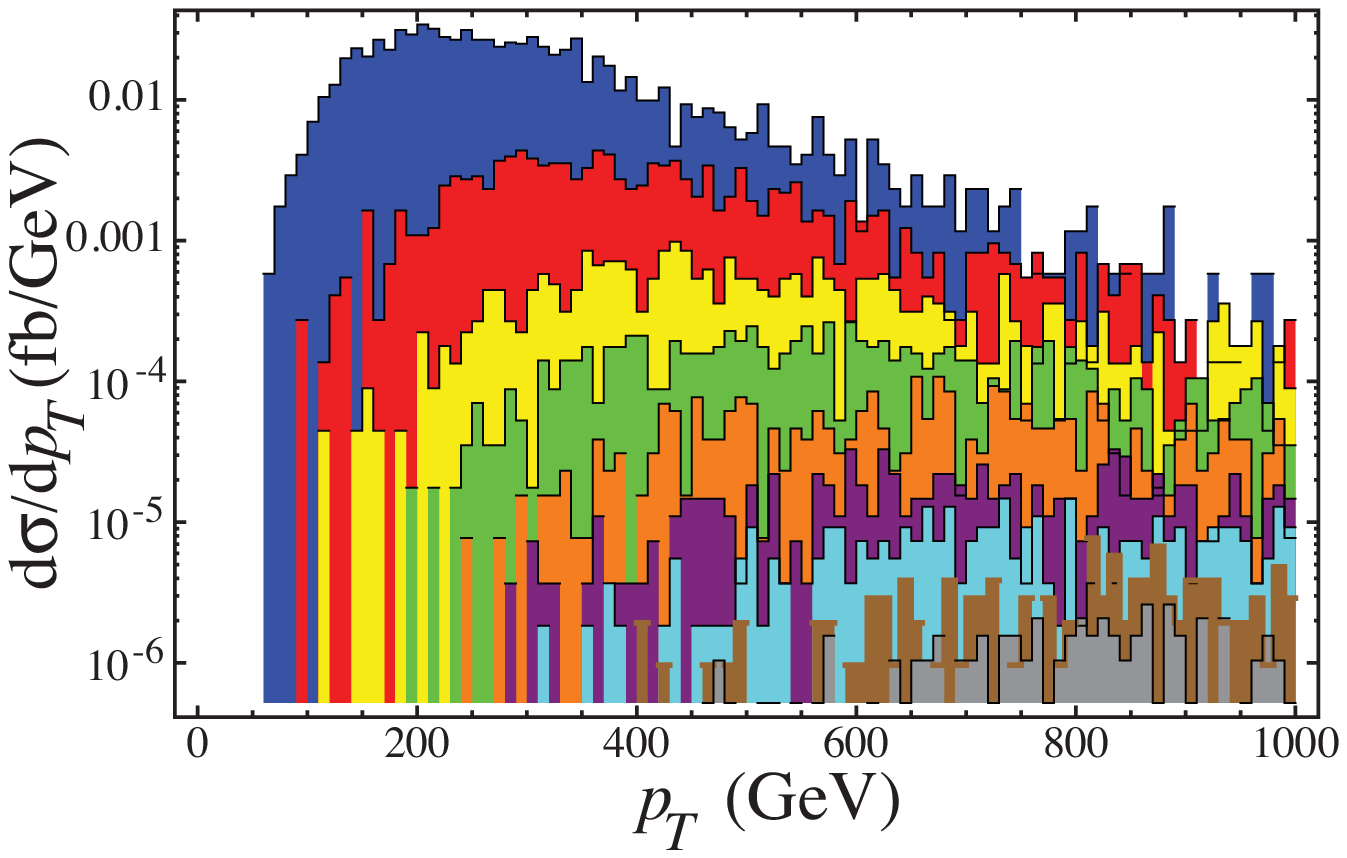}
\caption{Upper left: Maximum $|\eta|$ of $X^\pm_3$ from $X^+_3X^-_3+\rm{jets}$ events passing MET$>65$~GeV trigger as well as $X^\pm$ $|\eta|$ and isolation cut at $\sqrt{s} =13$~TeV. Lower left: Minimum $p_T$ of $X^\pm_3$from these events. Upper right: Minimum $|\eta|$ of $X^\pm_3$ from events that pass all cuts applied previous, plus the requirement that $Tc\tau >12$~cm for both $X^+_3$ and $X^-_3$. Lower right: Minimum $p_T$ of $X^\pm_3$ from those same events. Color coding denotes choice of $X_3^0$ mass and is the same as in Fig.~\ref{fig:tripletctau}. \label{fig:tripletX}}
\end{figure}

In Fig.~\ref{fig:tripletjets}, the jet $p_T$ and $|\eta|$ differential distributions for events both
before and after the $\mbox{min}[Tc\tau]>12$~cm cut are shown. Fig.~\ref{fig:tripletX} shows the differential
distribution of the minimum value of $p_T$ and maximum value of $|\eta|$ for the
$X^+_3X^-_3$ pair, again before and after the application of the $Tc\tau$ cut.
The effect of the $Tc\tau$ cut is (besides reducing the overall number of events) to push
the $\eta$ distribution to lower values and the $p_T$ distribution upward. This is as expected,
as the $Tc\tau$ constraint can more easily be satisfied by $X^\pm_3$ traveling in the central region.
The histogram bins have been weighted by total cross section and total number of generated events for each
mass choice. Note that small statistics become an issue after all cuts are applied.

It is difficult to determine the background rate for these stub
events. From a purely theoretical standpoint, there are no
irreducible backgrounds, as no SM particles have $c\tau$ lengths
that can compete with that of the triplets. However, there might be
a large background of low energy charged particles from the
underlying event, which will cause random hits in the inner detector
layers. These random hits could coincidentally line up in the pixel
tracker and register a fake event.

The number of such fake tracks will likely remain unknown without experimental measurement of the characteristics
of the underlying event. However, by exploiting the random nature of the background events, the
tracks resulting from the true triplets may be distinguished. We propose a few possible techniques
here.

The random nature of the fakes means that, even if two such false
tracks appear in a single event, it is highly unlikely that they will correctly point
back to the primary vertex (identified as the origin of the trigger jets).
High $p_T$ tracks, such as the ones we are interested
in, do not bend much inside the magnetic field of the detector. While this makes the
measurement of $p_T$ difficult, it does mean that the track vertex can be identified
to within $50\mu$m \cite{Cucciarelli:2006mt}. Since much of the background noise
in the event will presumably be due to pile-up, with up to 12 interactions separated
by 5~mm on average, this ability to unambiguously identify the primary vertex
should allow some subtraction of activity due to multiple interactions.

By requiring two stubs in each event, the number of fake events
passing our cuts can be reduced. If charge information for each
track can be obtained we would have an additional handle to identify
background, as the true triplet events must have one positive and
one negative particle. This may be aided by the requirement of 4-hit
or 5-hit tracks, since charge will be measured by the curvature of
the particle as it moves through the detector. With only three hits,
charge determination may be difficult or impossible. As mentioned,
long tracks come at a significant price in event rate. Also, we
found no study on the effectiveness of charge discrimination on
massive particles using only the inner tracking layers, making the
efficiency of this technique unknown.

Once we have a possible signal in the two-stub channel, we can
gain confidence that it is in fact new physics by considering the
single-stub events. In our simple models we can make precise
predictions of how many single track events with $Tc\tau>12$~cm
should be present compared to double-tracks. Naively, the ratio
should be about $4:1$ ($X^+X^0$, $X^-X^0$, and $X^+X^-$ where one
particle decays quickly). In reality, the ratio is skewed: among
other reasons, the $X^+X^0$ events are produced more copiously than
$X^-X^0$, as the LHC is a $p-p$ machine, not $p-\bar{p}$. From
simulation, we find approximately $5.5$ single track events for
every double track event in the triplet scenario. Additional
subtleties would arise if the new physics is some higher multiplet
of $SU(2)_L$, with several possible charged states. Observation of a
pattern would give us more reason to believe that the signal is in
fact a sign of new physics, rather than arising from background.

Additional confidence in our signal might be obtained by comparing the number of
single tracks passing
through four and five layers with the number passing through three.
Since this distribution should follow an exponential decay curve, it
may be possible to confirm that the signal seen in the pixel
detector is in fact not background.

These techniques, which all rely on the random nature of the
spurious track background, may provide tools which will determine
whether a sample of short tracks originates from new physics of the
type exemplified by the triplets. However, as stated previously, the
true background to the $X_3^+X_3^-$ tracks cannot be reliably
estimated without actual experimental data. Moreover the exponential
tail will have large fluctuations for small event numbers. Thus,
all these methods for background reduction are necessarily somewhat
speculative and we hope there will be a more rigorous study by the
two experimental collaborations.

We now turn to the discovery potential of the $X_2$ doublets. Recall
the doublets have larger mass splitting and are thus more difficult to detect.
Nonetheless,  the current bounds are so
weak that we can still place an interesting bound light doublets. At low
enough mass, the LHC will have a large enough production cross section
so that a sufficient number of double pairs on the exponential tail will make it
through all three layers.

The key requirement of the triplet discovery was the
$\mbox{min}[Tc\tau]>12$~cm for $X^\pm_3$.\footnote{Tracks passing
four or more layers are exponentially suppressed and have
essentially zero cross section for the doublet models. We shall
therefore only consider the three-layer tracks from this point on.}
A quick comparison of the decay
 length of the doublets $X^\pm_2$ with that of the triplets (as seen in Fig.~\ref{fig:ctau})
 makes it clear that such a strategy will be less promising for doublet searches. The difference in $c\tau$
 can be traced to the non-zero hypercharge of $X_2$, which leads to a larger mass splitting (Eq.~(\ref{eq:deltam})).

\begin{figure}[ht]

\includegraphics[width=0.5\columnwidth]{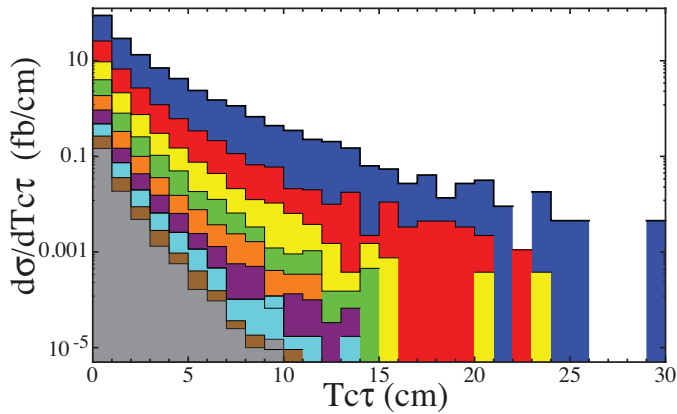}

\caption{Differential cross section versus transverse path length
$Tc\tau$ for $X^\pm_2$ from $X^+_2X^-_2+\rm{jets}$ events at $\sqrt{s}=13$~TeV,
triggering on $65$~GeV MET signature. Cuts of $\Delta R >0.4$ and $|\eta|<2.5$
have been applied, but no cut on $Tc\tau$ has been added (see text for more detail).
Color coding is as in Fig.~\ref{fig:tripletctau}. \label{fig:doubletctau}}
\end{figure}

In Fig.~\ref{fig:doubletctau}, the total differential cross section
for $X^\pm_2$ as a function of $Tc\tau$ is shown. As the
characteristic distance traveled is on the order of $1$~cm, the
cross section for events where both $X^+_2$ and $X^-_2$ survive a
$Tc\tau$ distance of $12$~cm is suppressed by a factor of approximately
$e^{-2\times 12/1}$ compared the production cross section. With this
suppression (compared to that of $e^{-2\times 12/10}$ for the
triplet case), almost no doublet events survive the application of
the $Tc\tau$ cuts. For the smallest choice of mass we explored in
the triplet case, $M=200$~GeV, we find only four events out of the
100,000 generated that passed all cuts, giving a nominal cross
section of $\sim 0.02$~fb for events with min$[Tc\tau]>12$~cm.

We therefore restrict our attention to a lower mass,  $100$~GeV.
This permits sufficient events to be seen for two reasons.  First,
the characteristic $c\tau$ increases with lower mass (see
Fig.~\ref{fig:ctau}), from $\sim 1$~cm when $M=200$~GeV to $1.8$~cm
when the mass is $100$~GeV. Second,  lower mass particles have a
much larger production cross section, increasing the probability
that some events will contain an $X^+_2X^-_2$ pair that both survive
long enough to hit the requisite three layers. After applying the
now-standard cuts on $\eta$, $\Delta R$, and min$[Tc\tau]>12$~cm,
we find a total cross section of $0.4$~fb. At CMS, the third layer of the
pixel is slightly closer to the beam, at 10~cm, rather than 12. While
this has a negligible effect for long life-time triplets, it amounts to
approximately one lifetime for the doublets. Using the CMS distance, we
require min$[Tc\tau]>10$~cm and find
a cross section of $0.9$~fb. From this we conclude that the LHC will  be able to probe
doublet models up to about $\sim 100$~GeV. While this is much worse
than the reach of triplet models, it nonetheless improves on the
current bound of $70$~GeV which can be obtained from LEP-II
\cite{Thomas:1998wy}.

The discovery potential is effectively set by the value of $\Delta
M$ (via its effect on the width and thus decay distance $Tc\tau$).
This accounts for the large difference in experimental sensitivity
between  triplets and doublets. In the next section we consider
relaxing the constraint between $M$ and $\Delta M$. This allows us
to explore interpolating cases  between the $\sim 150$~MeV splitting
of triplets to the $\sim 300$~MeV splittings of doublets, and
investigate the range of mass splittings which the LHC can probe, as
well as consider the possibility of an anomaly-mediated wino LSP
scenario.

\section{Perturbations on a Theme} \label{sec:complications}

If we assume that lepton number is an exact symmetry (or
alternatively, that the $X$ fields have their own global symmetry
that translates to a conserved quantum number), no tree-level terms
can be added to Eq.~(\ref{eq:lagrangian1}). Higher-dimensional terms
such as $HH X \bar{X}$ (where $H$ is the SM Higgs) are possible, but
do not cause an additional splitting between the neutral and charged
components.

However, a tree-level contribution to the splitting is possible when
 there is more than one new $SU(2)_L$ multiplet. Consider the case
when both $X_2$ and $X_3$ are present in the spectrum; with
tree-level masses $M_2$ and $M_3$ respectively. Then the Lagrangian
becomes
\begin{equation}
{\cal L} = i\bar{X}_2 \slashed{D} X_2 + i \bar{X}_2 \slashed{D} X_3
- M_2 \bar{X}_2X_2 - M_3 \bar{X}_3 X_3 + y H \bar{X}_2 X_3 +
\mbox{h.c.} \label{eq:lagrangian2}
\end{equation}
Once the Higgs field obtains its vacuum expectation value $v$, the
charged and neutral doublets and triplets mix. Assuming the
combination $yv$ is small compared to $M_2$ or $M_3$, the splitting
between the lowest charged and neutral eigenvalues is $yv$.

This effect can be realized in supersymmetry, which has vector
fermions in the form of the triplet winos $\tilde{W}$ and the
doublet higgsinos $\tilde{H}_u$ and $\tilde{H}_d$ (the presence of
two doublets is only a minor complication). As we are interested in
the phenomenology of a charged particle decaying into a stable
neutral state,  we only explore when the LSP $\tilde{\chi}^0_1$ and
NLSP $\tilde{\chi}^\pm_1$ are linear combinations of
$\tilde{W}^0/\tilde{H}^0$ and $\tilde{W}^\pm/\tilde{H}^\pm$
respectively. This can be arranged by setting the gaugino mass $M_2
< |M_1|,|\mu|$ (we also assume $M_2$ is real and positive). For
simplicity, we assume that all other supersymmetric masses are
large.

As a specific example,\footnote{This derivation
follows that of  \cite{Feng:1999fu}.} a wino LSP can be realized in
anomaly-mediated supersymmetry \cite{Randall:1998uk,Giudice:1998xp},
where the gaugino masses $M_i$ are
\begin{equation}
M_i = -b_i g_i^2 M_{\rm SUSY}. \label{eq:anommass}
\end{equation}
Here $M_{\rm SUSY}$ is the overall SUSY breaking scale, $i=1,2,3$ are the $U(1)_Y$,
$SU(2)_L$, and $SU(3)_C$ gauge groups (with couplings $g_i$) respectively, and $b_i$ are the one-loop
$\beta$ function coefficients. This leads to a mass ratio of $M_1: M_2:M_3 = 3.3:1:-10$.

The tree-level mass matrix for the neutralinos in the $(-i\tilde{B},-i\tilde{W}^3,\tilde{H}_1^0,\tilde{H}_2^0)$ is
\begin{equation}
{\bf M}_{\tilde{\chi}^0} = \left( \begin{array}{cccc} M_1 & 0 & -m_Z c_\beta s_W & m_Z s_\beta s_W \\ 0 & M_2 & m_Z c_\beta c_W & -m_Z s_\beta c_W \\ -m_Z c_\beta s_W & m_Z c_\beta c_W & 0 & -\mu \\ m_Z s_\beta s_W & -m_Z s_\beta c_W & - \mu & 0 \end{array}\right) \label{eq:neutralino}
\end{equation}
where $c_\beta = \cos \beta$, $s_\beta = \sin\beta$, $c_W = \cos\theta_W$ and $s_W = \sin\theta_W$. The
mass matrix for the charginos in the $(-i\tilde{W}^\pm,\tilde{H}^\pm)$ basis is
\begin{equation}
{\bf M}_{\tilde{\chi}^\pm} = \left(\begin{array}{cc} M_2 & \sqrt{2}m_W s_\beta \\ \sqrt{2} m_W c_\beta & \mu \end{array}\right) .\label{eq:chargino}
\end{equation}
Clearly, we could diagonalize ${\bf M}_{\tilde{\chi}^\pm}$ and ${\bf M}_{\tilde{\chi}^0}$ to
 find the mass difference $\Delta M_{\rm tree} \equiv m_{\tilde{\chi}^\pm_1}-  m_{\tilde{\chi}^0_1}$.
 In the limit of large $|\mu|$, a useful analytic expression can be obtained by expanding the mass difference in terms of
 inverse
  powers of $\mu$. We find that the $\mu^{-1}$ term vanishes, and the first non-zero term is
\begin{equation}
\Delta M_{\rm tree} =\left(M_2(1-\sin 2 \beta)+\frac{m_W^2 \tan^2\theta_W}{(M_1-M_2)}\sin^22\beta\right) \frac{m_W^2}{\mu^2}\label{eq:deltatree}.
\end{equation}
This is in addition to the loop splitting of Eq.~(\ref{eq:deltam}).
We see that, if $M_1$ and $M_2$ are ${\cal O}(100~\mbox{GeV})$ while
$|\mu|$ is ${\cal O}(\mbox{TeV})$, then $\Delta M_{\rm tree}$ is of
the same order as the loop contribution: a few hundred MeV.

By appropriate choices of the magnitudes and signs of $M_1$, $M_2$,
$\mu$, and $\tan\beta$ (perhaps requiring us to relax the anomaly-mediated
predictions for the relative size of $M_1$ and $M_2$) we
can effectively dial the total mass splitting. This sets the scale
of $Tc\tau$ and, as seen in Section \ref{sec:triplet}, controls
whether a particular model can be discovered at the LHC.

If we strictly abide by the gaugino mass ratio of $3.3 : 1: -10$ in
anomaly mediated supersymmetry breaking, then from
Eq.~(\ref{eq:deltatree}), we see that the tree level splitting can
only be positive. Added to the triplet loop-induced splitting, this
 means that the pion channel is
 always open to the NLSP. In order to close this channel,
 the tree contribution would have to be negative, requiring $M_2 > M_1$, which
 does not occur in a conventional anomaly-mediated spectrum.

To explore the dependence of an LHC discovery on $\Delta M$, we
consider SUSY spectra with a fixed mass of the neutralino LSP
$\tilde{\chi}_1^0$ and vary the mass of the NLSP
$\tilde{\chi}^\pm_1$. To discuss experimental reach, we first  treat
$\Delta M$ as a free parameter. Afterwards we will consider what
ranges of anomaly-mediated input parameters  that the LHC can
access.

The full spectrum was calculated using SuSpect
2.41\cite{Djouadi:2002ze}, with input parameters tuned to give the required splittings.
 As we are taking our cue from anomaly mediated scenarios with
the hierarchy $|M_2| < |M_1| \ll |\mu|$, the sfermions tend to be
very massive ($\gtrsim 2$~TeV). This is not always true, for the
lightest neutralinos with masses near the current experimental
bounds ($\sim 100$~GeV \cite{Abdallah:2003gv}) the squark masses in
an anomaly mediated spectrum must be on the order of $500$~GeV for
$\Delta M \sim 200$~MeV, and $1500$~GeV for splittings of $350$~MeV.
However, in all cases we investigated, the sfermions decouple from
the low energy theory and do not greatly affect the production cross
section. Even in the previous examples, changing the lightest squark
mass from $500$~GeV to $1500$~GeV while keeping the LSP/NLSP masses
at $100$~GeV increases the total production cross section by only
$2\%$.

We set the LSP mass $m_{\tilde{\chi}_1^0}$ equal to $100$, $200$,
$300$, and $400$~GeV. In all cases, we generate $100,000$
$\tilde{\chi}^+_1\tilde{\chi}^-_1+\mbox{jets}$ events in MadGraph
at $\sqrt{s} =13$~TeV. After event generation, we vary the mass
splitting $\Delta M$ from $145$ to $220$~MeV. Such change has no
appreciable effect on the kinematics of an individual event with the
exceptions of the energy of the light decay products and the decay
length. The latter, of course, is the parameter of interest in our
study while the former can be ignored, as the pions cannot be used
in track identification. The key point here is that, if $\Delta M$
is the only parameter that is adjusted, any event that passed
the acceptance cuts ({\it i.e.}~MET trigger, $\Delta R$, $|\eta|$)
will continue to do so; changing $\Delta M$ affects only whether a
particular event will pass the $Tc\tau$ cuts.

\begin{figure}[ht]

\includegraphics[width=0.5\columnwidth]{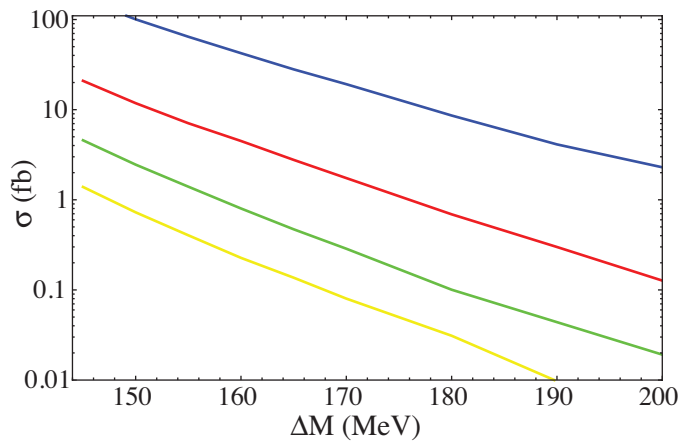}
\caption{Cross section of $\tilde{\chi}^+_1\tilde{\chi}^-_1+\mbox{jets}$ passing MET$>65$~GeV trigger
as well as $|\eta|$ and isolation cut at $\sqrt{s} =13$~TeV as a function of mass splitting
$\Delta M \equiv m_{\tilde{\chi}_\pm^0} - m_{\tilde{\chi}_1^0}$.
The blue lines corresponds to a neutralino
mass $m_{\tilde{\chi}_1^0} =100~$GeV, while red is $200$~GeV, green $300$~GeV, and yellow $400$~GeV.
\label{fig:splittingsigma}}
\end{figure}

After selecting a mass difference, we then apply a MET$>65$~GeV trigger, as well as the
$\eta$, jet isolation, and $Tc\tau$ cuts as outlined in
Section~\ref{sec:triplet}.
The resulting cross sections of events
with two charged tracks with min$[Tc\tau]>12~$cm are displayed in
Fig.~\ref{fig:splittingsigma}.

Assuming $10(100)$~fb$^{-1}$ of luminosity, we find  that a 400~GeV
NLSP can  be discovered only if the splitting is less than
$\sim155(175)$~MeV (again, assuming 5 events are required). A 100
GeV NLSP can be discovered if the splitting is larger, on the order
of $220~$MeV for the lower luminosity, and perhaps as high as
$240$~MeV for 100~fb$^{-1}$. Due to the exponential fall off on
the numbers of events as we go to larger splittings, small
statistics become an issue.

From these results, we
can also determine the range of anomaly-mediation parameters that
the LHC should  probe. The loop-induced mass splitting for a
$100$~GeV triplet is $145$~MeV,  meaning that the LHC will be
sensitive to a tree-level splitting of $75$~MeV. From the splitting
between the lightest eigenvalues of the neutralino and chargino mass
matrices (Eqs.~(\ref{eq:neutralino})  and (\ref{eq:chargino})), we
see this splitting corresponds to $|\mu| \gtrsim 5$~TeV for $\tan
\beta =10$. For a $400$~GeV neutralino, the LHC would be sensitive a
tree-level splitting less than $15$~MeV (at high luminosity), which
requires $|\mu | \gtrsim 6$~TeV.

Assuming a universal scalar mass $m_0$, then the wino mass $M_2$ can
be related (via the anomaly mediation mass $M_{\rm aux}$) to the $\mu$
parameter and $m_0$ by the requirements of EWSB. For $m_0 \lesssim 2$~TeV
and wino masses of ${\cal O}(100~\mbox{GeV})$, the required $|\mu|$ value tends to
be much less than the $5-6$~TeV required for sufficiently small $\Delta M$.
As a specific example, for $\mbox{sgn}(\mu)<0$ (preferred from $b\to s\gamma$
for large $\tan\beta$), $\tan \beta \gtrsim 10$, $m_0 < 2$~TeV, and a wino mass of $400$~GeV
 corresponds to $\mu \sim -2500$~GeV \cite{Feng:1999hg}. At the lower end of the wino mass scale,
$M_2 = 100$~GeV corresponds to $\mu \sim -600$~GeV for the same parameters.

When $m_0$ is relatively small leading to a more natural spectrum,
the stub signature for anomaly mediated degenerate winos is
between a rock and a hard place. If $|\mu|$ is small enough to allow
for a light NLSP (and thus a large cross section at the LHC), the
splitting will be large, and the NLSP will decay rapidly and
invisibly to the LSP. On the other hand, if the $\mu$ parameter is
large, the chargino's $c\tau$ will be big, but the overall rate of
production will be too small for meaningful statistics to be
collected.

In order to have $|\mu|\sim 5$~TeV (and thus the requisite small
$\Delta M$), with the wino spectrum sufficiently light to be visible
[$m_0$ would need to also be around $5$~TeV. This would create
an unnatural spectrum, with heavy scalars of ${\cal
O}(5~\mbox{TeV})$ and light gaugino masses of a few hundred GeV.

One can have satisfactory and visible anomaly-mediated-based models
when they are supplemented by a non-universal scalar mass. Relaxing
the requirement of universal $m_0$ masses would allow the squarks
and sleptons to remain light although the Higgs mass would still be
unnaturally big.

If we abandon  anomaly-mediation altogether, then we are no longer
bound by the relation between $M$ and $\Delta M$ shown in
Eq.~(\ref{eq:deltatree}), and we may take our results for maximum
$\Delta M$ as a function of $M$ as
the expected reach of the LHC for degenerate chargino/neutralino
pairs. Implicit in this statement is the assumption that the squarks
and sleptons are much more massive than the winos. If they are not
then the overall production cross sections must be adjusted
accordingly.

Additionally, if $M_2$ is no longer required to be larger than
$M_1$, there is the possibility that the tree level splitting is
negative. In this case, in addition to the splittings investigated
above (which are all larger than the loop-level contribution), we
have the possibility of splittings smaller than the pion mass. In
such cases, detection would be straightforward. With $\Delta M <
m_\pi$, and the decay proceeds through the $\tilde{\chi}_1^0 \ell
\nu_\ell$ three-body final state. From Fig.~\ref{fig:ctau}, we see
that the charginos will travel on the order of several meters before
decaying.  Stubs would no longer be the signal for this model.

To demonstrate this possibility, we consider an LSP mass of
$400$~GeV and $\Delta M = 100$~MeV. With a MET$>65~$GeV cut, the
production cross section for
$\tilde{\chi}^+_1\tilde{\chi}^-_1+\mbox{jets}$ events passing the
$|\eta|$ and isolation cuts is $10.0$~fb. Of these events, $8.1$~fb
have both $\tilde{\chi}_1^\pm$ states decay after the requisite
$12$~cm $Tc\tau$ cut, and $0.04$~fb after $10$~meters (the
characteristic distance to the muon system at ATLAS). The total
cross section for metastable charged particles at the LHC will in
fact be larger, since it will be possible to trigger on the heavy
particles themselves, rather than the associated jet production.
These events would show up in the muon system as heavy, slow
`muons,' essentially identical to the expected signature of new stable charged
particles.
The current lower bound on the mass of wino-like stable LSP/NLSP
pairs is $206$~GeV from D0 \cite{Abazov:2008qu}, expected to rise to
$\sim 600$~GeV at the LHC (at $\sqrt{s} = 14~$TeV)
\cite{Raklev:2009mg}. As the searches for stable charged particles
differ significantly from the stub searches required for unstable
particles, we do not consider this scenario further here.  However,
we do note that the presence and $Tc\tau$ distribution of charged
tracks which disappear in flight may provide a method of determining
the lifetime and mass splittings if the `stable' charged particles
are in fact merely a long-lived (on detector timescales) NLSP.

\section{Conclusions} \label{sec:conclusions}

New weakly charged multiplets with nearly degenerate masses are an
excellent example of non-standard new physics at the TeV-scale that
might be produced at the LHC yet evade observation with current
search strategies. Though not required by most solutions to the
hierarchy, naturalness, and dark matter problems, they  might exist
and could be within observable reach.  Search strategies for such
particles certainly merit consideration.

In the simple models considered in this paper, {\it i.e.}~the
minimal examples with $SU(2)_L$ doublets or triplets, the new
multiplets do not have any strongly-coupled production mechanism.
Furthermore, their small mass splittings lead to very low energy SM
decay products in addition to the invisible neutral state. While the
anomaly-mediated models do contain strongly coupled states (gluinos
and squarks), these particles could be too heavy to be produced in
large numbers at the LHC.

Discovery of these models through standard techniques is impossible.
The weak production cross section means that the contribution to
jets+$\slashed{E}_T$ would be swamped by SM processes, and the low
energy decay products are completely lost in the noisy underlying
event. Note that larger representations of $SU(2)_L$ face the same
issues. Without much higher luminosity or energy, we don't expect
even representations to be visible since their width (for a given
mass) should be at least as large as in the doublet case.  For odd
multiplets with zero hypercharge, the mass difference between the
lowest charged and neutral states is the same as for the triplet,
but both production and decay widths will be bigger (due to the
larger Casimir) so the searches will not be manifestly simpler, and
in fact may have more in common with the doublet searches than the
triplets.

In this paper we discussed a potentially promising approach for
discovering  vector  multiplets at the LHC with nearly degenerate
elements that focuses on the short charged tracks formed by the
charged component. Though standard track-reconstruction techniques
require hits in all layers of the inner detector which
correlate with energy deposition in the calorimeters or activity in
the muon chamber, it should be possible to relax these constraints.
We suggest that the nearly degenerate  particles could be found by
 the ATLAS and CMS collaborations using the high resolution of the
inner pixel detector in order to find `stub' tracks passing through
the first three detector layers, using a missing $E_T$ trigger to
write the event to tape. With off-line analysis, one should be able
to identify events containing two isolated, charged high $p_T$
stubs.

Background rejection will be a major issue, and the rate of fake
events is difficult to estimate without experimental data. We expect
that the probability of two short tracks being formed by random hits
is very small, but at the present, this is mere conjecture. Whatever
the rate, requiring that the tracks and jets share a common primary
vertex should help reduce background. Requiring unlike charges
on the tracks would also help, though this measurement is expected
to be very difficult due to the stiffness of the high $p_T$ particles.
If necessary, stubs passing through four
or more layers could be used, although this will come
at a significant sacrifice of event rate.

Because of the dependence of particle lifetime on mass splitting in
the multiplet, the requirement that the charged particle pass at
least three layers of the pixel tracker makes the mass splitting
$\Delta M$ the linchpin of the discovery potential.   For $\Delta M
\sim 150$~MeV, as in the case of triplets, the new multiplets can be
seen for  masses below $350(550)$~GeV with $10(100)$~fb$^{-1}$ of data.
This reach drops precipitously with increasing $\Delta M$; $SU(2)_L$
doublets have splittings on the order of $300$~MeV, and for masses much
greater than $100$~GeV they are invisible at the LHC.
Larger splittings are likewise undetectable, until multi-particle
modes open at around $\sim 1$~GeV. However, the background for
particles that decay into  low energy multi-hadron states with large
impact parameters is unknown, and we have not considered such
searches in this paper. Of course, for $\Delta M < m_\pi$, the
lifetime is long enough to pass through all of the central tracker
and even into the calorimeters and muon system. Discovery is then
relatively straightforward, subject to the same limitations as searches
for new stable charge particles.

If a signal were found in the stub search at the LHC, it will be
difficult to determine the underlying theory.
For example, scalar triplets \cite{Ross1975,Gunion:1989ci,Blank:1997qa,
Forshaw:2001xq,Forshaw:2003kh,Chen:2006pb,Chankowski:2006hs}, which could be a low-energy remnant of
grand unified models that avoid rapid proton decay \cite{Dorsner:2005fq}, as well as relaxing constraints from
electro-weak precision data, might yield similar signals to fermionic triplets.
As with the fermions we considered above, the charged scalar can be seen only when the splitting
from the neutral state is close to $m_\pi$ and triplet's mass is on the order of a few hundred GeV, which provides
a sufficient rate for detection. In both cases, the observed signal would be very similar: a charged stub in the
central tracker disappearing with no energy deposition in the outer detector.

However, unlike the fermions, scalar triplets can have tree-level
couplings with the SM Higgs boson. Generically, these couplings will
lead to significant changes in the branching ratio of the neutral
Higgs decaying to photons \cite{FileviezPerez:2008bj}. Measurement
of this rate, combined with the observation of stub events, could
allow the boson and fermion models to be distinguished.

Anomaly-mediated supersymmetric theories generally contain nearly
degenerate neutralinos and charginos.    We have seen that the reach
at the LHC for such models is greater than $400$~GeV for splittings
on par with the pion mass, down to $\sim 100$~GeV neutralinos for
splittings around $250$~MeV. However since anomaly-mediated
spectra tend to have the LSP mass increase with $|\mu|$ while the
splitting decreases,  natural models of anomaly-mediated
supersymmetry will in general not contain an LSP/NLSP pair with
small enough splitting to provide a sufficiently large lifetime
while also having a small enough mass to be produced copiously at
the collider.

\section*{Acknowledgements}

The authors would like to thank Massimiliano Chiorboli, Beate
Heinemann,  Dorian Kcira,  Greg Landsman, Vladimir Litvine, Maurizio
Pierini,
 Michael Schmitt, Sezen Sekmen, and Maria Spiropulu
for their invaluable expertise and advice. LR thanks Caltech for a
Moore Fellowship and their hospitality and NYU and the CCPP for its
hospitality while some of this work was completed. We would also
like to thank the Aspen Center for Physics for providing a
simulating and productive environment for discussion and work.


\begin{thebibliography}{99}

%\cite{Wess:1974tw}
\bibitem{Wess:1974tw}
  J.~Wess and B.~Zumino,
  %``Supergauge Transformations in Four-Dimensions,''
  Nucl.\ Phys.\  B {\bf 70}, 39 (1974).
  %%CITATION = NUPHA,B70,39;%%

%\cite{ArkaniHamed:1998nn}
\bibitem{ArkaniHamed:1998nn}
  N.~Arkani-Hamed, S.~Dimopoulos and G.~R.~Dvali,
  %``Phenomenology, astrophysics and cosmology of theories with  sub-millimeter
  %dimensions and TeV scale quantum gravity,''
  Phys.\ Rev.\  D {\bf 59}, 086004 (1999)
  [arXiv:hep-ph/9807344].
  %%CITATION = PHRVA,D59,086004;%%

%\cite{Randall:1999ee}
\bibitem{Randall:1999ee}
  L.~Randall and R.~Sundrum,
  %``A large mass hierarchy from a small extra dimension,''
  Phys.\ Rev.\ Lett.\  {\bf 83}, 3370 (1999)
  [arXiv:hep-ph/9905221].
  %%CITATION = PRLTA,83,3370;%%

 %\cite{Weinstein:1973gj}
\bibitem{Weinstein:1973gj}
  M.~Weinstein,
  %``Conserved Currents, Their Commutators And The Symmetry Structure Of
  %Renormalizable Theories Of Electromagnetic, Weak And Strong Interactions,''
  Phys.\ Rev.\  D {\bf 8}, 2511 (1973).
  %%CITATION = PHRVA,D8,2511;%%

%\cite{Weinberg:1979bn}
\bibitem{Weinberg:1979bn}
  S.~Weinberg,
  %``Implications Of Dynamical Symmetry Breaking: An Addendum,''
  Phys.\ Rev.\  D {\bf 19}, 1277 (1979).
  %%CITATION = PHRVA,D19,1277;%%

%\cite{Susskind:1979}
\bibitem{Susskind:1979}
 L.~Susskind,
 %``Dynamics of spontaneous symmetry breaking in the Weinberg-Salam theory,''
 Phys.\ Rev.\ D {\bf 20}, 2619 (1979).

 %\cite{Moroi:1999zb}
\bibitem{Moroi:1999zb}
  T.~Moroi and L.~Randall,
  %``Wino cold dark matter from anomaly-mediated SUSY breaking,''
  Nucl.\ Phys.\  B {\bf 570}, 455 (2000)
  [arXiv:hep-ph/9906527].
  %%CITATION = NUPHA,B570,455;%%

%\cite{Cirelli:2005uq}
\bibitem{Cirelli:2005uq}
  M.~Cirelli, N.~Fornengo and A.~Strumia,
  %``Minimal dark matter,''
  Nucl.\ Phys.\  B {\bf 753}, 178 (2006)
  [arXiv:hep-ph/0512090].
  %%CITATION = NUPHA,B753,178;%%

%\cite{Cui:2009xq}
\bibitem{Cui:2009xq}
  Y.~Cui, D.~E.~Morrissey, D.~Poland and L.~Randall,
  %``Candidates for Inelastic Dark Matter,''
  JHEP {\bf 0905}, 076 (2009)
  [arXiv:0901.0557 [hep-ph]].
  %%CITATION = JHEPA,0905,076;%%

%\cite{FileviezPerez:2008bj}
\bibitem{FileviezPerez:2008bj}
  P.~Fileviez Perez, H.~H.~Patel, M.~J.~Ramsey-Musolf and K.~Wang,
  %``Triplet Scalars and Dark Matter at the LHC,''
  Phys.\ Rev.\  D {\bf 79}, 055024 (2009)
  [arXiv:0811.3957 [hep-ph]].
  %%CITATION = PHRVA,D79,055024;%%

%\cite{Drees:1990yw}
\bibitem{Drees:1990yw}
  M.~Drees and X.~Tata,
  %``Signals for heavy exotics at hadron colliders and supercolliders,''
  Phys.\ Lett.\  B {\bf 252}, 695 (1990).
  %%CITATION = PHLTA,B252,695;%%

%\cite{Aaltonen:2009ke}
\bibitem{Aaltonen:2009ke}
  T.~Aaltonen {\it et al.}  [CDF Collaboration],
  %``Search for Long-Lived Massive Charged Particles in 1.96 TeV $\bar{p}p$
  %Collisions,''
  arXiv:0902.1266 [hep-ex].
  %%CITATION = ARXIV:0902.1266;%%

%\cite{Feng:1999fu}
\bibitem{Feng:1999fu}
  J.~L.~Feng, T.~Moroi, L.~Randall, M.~Strassler and S.~f.~Su,
  %``Discovering supersymmetry at the Tevatron in Wino LSP scenarios,''
  Phys.\ Rev.\ Lett.\  {\bf 83}, 1731 (1999)
  [arXiv:hep-ph/9904250].
  %%CITATION = PRLTA,83,1731;%%

%\cite{Gunion:1999jr}
\bibitem{Gunion:1999jr}
  J.~F.~Gunion and S.~Mrenna,
  %``A study of SUSY signatures at the Tevatron in models with near mass
  %degeneracy of the lightest chargino and neutralino,''
  Phys.\ Rev.\  D {\bf 62}, 015002 (2000)
  [arXiv:hep-ph/9906270].
  %%CITATION = PHRVA,D62,015002;%%

  %\cite{Ibe:2006de}
\bibitem{Ibe:2006de}
  M.~Ibe, T.~Moroi and T.~T.~Yanagida,
  %``Possible signals of Wino LSP at the Large Hadron Collider,''
  Phys.\ Lett.\  B {\bf 644}, 355 (2007)
  [arXiv:hep-ph/0610277].
  %%CITATION = PHLTA,B644,355;%%

  %\cite{Abreu:1999qr}
\bibitem{Abreu:1999qr}
  P.~Abreu {\it et al.}  [DELPHI Collaboration],
  %``Search for charginos nearly mass-degenerate with the lightest
  %neutralino,''
  Eur.\ Phys.\ J.\  C {\bf 11}, 1 (1999)
  [arXiv:hep-ex/9903071].
  %%CITATION = EPHJA,C11,1;%%

%\cite{Abbiendi:2002vz}
\bibitem{Abbiendi:2002vz}
  G.~Abbiendi {\it et al.}  [OPAL Collaboration],
  %``Search for nearly mass-degenerate charginos and neutralinos at LEP,''
  Eur.\ Phys.\ J.\  C {\bf 29}, 479 (2003)
  [arXiv:hep-ex/0210043].
  %%CITATION = EPHJA,C29,479;%%


%\cite{Randall:1998uk}
\bibitem{Randall:1998uk}
  L.~Randall and R.~Sundrum,
  %``Out of this world supersymmetry breaking,''
  Nucl.\ Phys.\  B {\bf 557}, 79 (1999)
  [arXiv:hep-th/9810155].
  %%CITATION = NUPHA,B557,79;%%

%\cite{Giudice:1998xp}
\bibitem{Giudice:1998xp}
  G.~F.~Giudice, M.~A.~Luty, H.~Murayama and R.~Rattazzi,
  %``Gaugino Mass without Singlets,''
  JHEP {\bf 9812}, 027 (1998)
  [arXiv:hep-ph/9810442].
  %%CITATION = JHEPA,9812,027;%%

%\cite{Abdallah:2003gv}
\bibitem{Abdallah:2003gv}
  J.~Abdallah {\it et al.}  [DELPHI Collaboration],
  %``Search for SUSY in the AMSB scenario with the DELPHI detector,''
  Eur.\ Phys.\ J.\  C {\bf 34}, 145 (2004)
  [arXiv:hep-ex/0403047].
  %%CITATION = EPHJA,C34,145;%%

%\cite{Decamp:1989fr}
\bibitem{Decamp:1989fr}
  D.~Decamp {\it et al.}  [ALEPH Collaboration],
  %``A PRECISE DETERMINATION OF THE NUMBER OF FAMILIES WITH LIGHT NEUTRINOS AND
  %OF THE Z BOSON PARTIAL WIDTHS,''
  Phys.\ Lett.\  B {\bf 235}, 399 (1990).
  %%CITATION = PHLTA,B235,399;%%
%\cite{Akrawy:1990zy}
%\bibitem{Akrawy:1990zy}
  M.~Z.~Akrawy {\it et al.}  [OPAL Collaboration],
  %``A measurement of the Z0 invisible width by single photon counting,''
  Z.\ Phys.\  C {\bf 50}, 373 (1991).
  %%CITATION = ZEPYA,C50,373;%%
%\cite{Aarnio:1990qe}
%\bibitem{Aarnio:1990qe}
  P.~A.~Aarnio {\it et al.}  [DELPHI Collaboration],
  %``STUDY OF THE LEPTONIC DECAYS OF THE Z0 BOSON,''
  Phys.\ Lett.\  B {\bf 241}, 425 (1990).
  %%CITATION = PHLTA,B241,425;%%

  %\cite{Thomas:1998wy}
\bibitem{Thomas:1998wy}
  S.~D.~Thomas and J.~D.~Wells,
  %``Phenomenology of Massive Vectorlike Doublet Leptons,''
  Phys.\ Rev.\ Lett.\  {\bf 81}, 34 (1998)
  [arXiv:hep-ph/9804359].
  %%CITATION = PRLTA,81,34;%%

%\cite{Chen:1995yu}
\bibitem{Chen:1995yu}
  C.~H.~Chen, M.~Drees and J.~F.~Gunion,
  %``Searching for Invisible and Almost Invisible Particles at $e~+e~-$
  %Colliders,''
  Phys.\ Rev.\ Lett.\  {\bf 76}, 2002 (1996)
  [arXiv:hep-ph/9512230].
  %%CITATION = PRLTA,76,2002;%%

%\cite{PausTalk}
\bibitem{PausTalk}
  C.~Paus [on behalf of the CMS collaboration],
  %`Trigger Strategies and Early Physics at CMS'
  %\href{http://indico.cern.ch/materialDisplay.py?contribId=10&materialId=slides&confId=55994}
  {\it Prepared for Berkeley Workshop on Physics Opportunities with Early LHC Data, Berkeley, USA, 6-8 May, 2009.}

%\cite{Bayatian:2006zz}
\bibitem{Bayatian:2006zz}
  G.~L.~Bayatian {\it et al.}  [CMS Collaboration],
  %``CMS physics: Technical design report,''
  %%CITATION = CMS-TDR-008-1;%%

%\cite{Aad:2009wy}
\bibitem{Aad:2009wy}
  G.~Aad {\it et al.}  [The ATLAS Collaboration],
  %``Expected Performance of the ATLAS Experiment - Detector, Trigger and
  %Physics,''
  arXiv:0901.0512 [hep-ex].
  %%CITATION = ARXIV:0901.0512;%%

%\cite{Adam:2005zf}
\bibitem{Adam:2005zf}
  W.~Adam {\it et al.}  [CMS Trigger and Data Acquisition Group],
  %``The CMS high level trigger,''
  Eur.\ Phys.\ J.\  C {\bf 46}, 605 (2006)
  [arXiv:hep-ex/0512077].
  %%CITATION = EPHJA,C46,605;%%

%\cite{Stelzer:1994ta}
\bibitem{Stelzer:1994ta}
  T.~Stelzer and W.~F.~Long,
  %``Automatic generation of tree level helicity amplitudes,''
  Comput.\ Phys.\ Commun.\  {\bf 81}, 357 (1994)
  [arXiv:hep-ph/9401258].
  %%CITATION = CPHCB,81,357;%%

%\cite{Sjostrand:2006za}
\bibitem{Sjostrand:2006za}
  T.~Sjostrand, S.~Mrenna and P.~Skands,
  %``PYTHIA 6.4 Physics and Manual,''
  JHEP {\bf 0605}, 026 (2006)
  [arXiv:hep-ph/0603175].
  %%CITATION = JHEPA,0605,026;%%

%\cite{Cucciarelli:2006mt}
\bibitem{Cucciarelli:2006mt}
  S.~Cucciarelli, M.~Konecki, D.~Kotlinski and T.~Todorov,
  %``Track reconstruction, primary vertex finding and seed generation with  the
  %pixel detector,''
  %%CITATION = CERN-CMS-NOTE-2006-026;%%

%\cite{Djouadi:2002ze}
\bibitem{Djouadi:2002ze}
  A.~Djouadi, J.~L.~Kneur and G.~Moultaka,
  %``SuSpect: A Fortran code for the supersymmetric and Higgs particle spectrum
  %in the MSSM,''
  Comput.\ Phys.\ Commun.\  {\bf 176}, 426 (2007)
  [arXiv:hep-ph/0211331].
  %%CITATION = CPHCB,176,426;%%

%\cite{Feng:1999hg}
\bibitem{Feng:1999hg}
  J.~L.~Feng and T.~Moroi,
  %``Supernatural supersymmetry: Phenomenological implications of
  %anomaly-mediated supersymmetry breaking,''
  Phys.\ Rev.\  D {\bf 61}, 095004 (2000)
  [arXiv:hep-ph/9907319].
  %%CITATION = PHRVA,D61,095004;%%

%\cite{Abazov:2008qu}
\bibitem{Abazov:2008qu}
  V.~M.~Abazov {\it et al.}  [D0 Collaboration],
  %``Search for Long-Lived Charged Massive Particles with the D0 Detector,''
  Phys.\ Rev.\ Lett.\  {\bf 102}, 161802 (2009)
  [arXiv:0809.4472 [hep-ex]].
  %%CITATION = PRLTA,102,161802;%%

\bibitem{Raklev:2009mg}
  A.~R.~Raklev,
  %``Massive Metastable Charged (S)Particles at the LHC,''
  arXiv:0908.0315 [hep-ph].
  %%CITATION = ARXIV:0908.0315;%%

%\cite{Ross1975}
\bibitem{Ross1975}
  D.A.~Ross, M.J.G.~Veltman,
  ``Neutral Currents and The Higgs Mechanism,'' Nucl.~Phys.~B 95:135,1975.

%\cite{Gunion:1989ci}
\bibitem{Gunion:1989ci}
  J.~F.~Gunion, R.~Vega and J.~Wudka,
  %``Higgs triplets in the standard model,''
  Phys.\ Rev.\  D {\bf 42}, 1673 (1990).
  %%CITATION = PHRVA,D42,1673;%%

%\cite{Forshaw:2001xq}
\bibitem{Forshaw:2001xq}
  J.~R.~Forshaw, D.~A.~Ross and B.~E.~White,
  %``Higgs mass bounds in a triplet model,''
  JHEP {\bf 0110}, 007 (2001)
  [arXiv:hep-ph/0107232].
  %%CITATION = JHEPA,0110,007;%%

%\cite{Forshaw:2003kh}
\bibitem{Forshaw:2003kh}
  J.~R.~Forshaw, A.~Sabio Vera and B.~E.~White,
  %``Mass bounds in a model with a triplet Higgs,''
  JHEP {\bf 0306}, 059 (2003)
  [arXiv:hep-ph/0302256].
  %%CITATION = JHEPA,0306,059;%%

%\cite{Chen:2006pb}
\bibitem{Chen:2006pb}
  M.~C.~Chen, S.~Dawson and T.~Krupovnickas,
  %``Higgs triplets and limits from precision measurements,''
  Phys.\ Rev.\  D {\bf 74}, 035001 (2006)
  [arXiv:hep-ph/0604102].
  %%CITATION = PHRVA,D74,035001;%%

%\cite{Chankowski:2006hs}
\bibitem{Chankowski:2006hs}
  P.~H.~Chankowski, S.~Pokorski and J.~Wagner,
  %``(Non)decoupling of the Higgs triplet effects,''
  Eur.\ Phys.\ J.\  C {\bf 50}, 919 (2007)
  [arXiv:hep-ph/0605302].
  %%CITATION = EPHJA,C50,919;%%

%\cite{Blank:1997qa}
\bibitem{Blank:1997qa}
  T.~Blank and W.~Hollik,
  %``Precision observables in SU(2) x U(1) models with an additional Higgs
  %triplet,''
  Nucl.\ Phys.\  B {\bf 514}, 113 (1998)
  [arXiv:hep-ph/9703392].
  %%CITATION = NUPHA,B514,113;%%

%\cite{Dorsner:2005fq}
\bibitem{Dorsner:2005fq}
  I.~Dorsner and P.~Fileviez Perez,
  %``Unification without supersymmetry: Neutrino mass, proton decay and  light
  %leptoquarks,''
  Nucl.\ Phys.\  B {\bf 723}, 53 (2005)
  [arXiv:hep-ph/0504276].
  %%CITATION = NUPHA,B723,53;%%

\end{thebibliography}
\end{document}